# Some Applications of Coding Theory in Computational Complexity

Luca Trevisan[*]

May 20, 2004

**Abstract**

Error-correcting codes and related combinatorial constructs play an important role in several recent (and old) results in computational complexity theory. In this paper we survey results on locally-testable and locally-decodable error-correcting codes, and their applications to complexity theory and to cryptography.

Locally decodable codes are error-correcting codes with sub-linear time error-correcting algorithms. They are related to private information retrieval (a type of cryptographic protocol), and they are used in average-case complexity and to construct "hard-core predicates" for one-way permutations. Locally testable codes are error-correcting codes with sub-linear time error-detection algorithms, and they are the combinatorial core of probabilistically checkable proofs.

## Contents

[*]luca@cs.berkeley.edu. U.C. Berkeley, Computer Science Division. Work supported by NSF grant CCR-9984703, a Sloan Research Fellowship, and an Okawa Foundation Grant.

# 1 Introduction

Many recent (and not-so-recent) results in complexity theory rely on error-correcting codes. The use of coding-theoretic concepts, constructions and algorithms has been a major theme of complexity theoretic results in the past few years, and so has been the re-interpretation of older results in a coding-theoretic language.

An error-correcting code is a mapping $C : \{0,1\}^k \to \{0,1\}^n$ (or, more generally, $C : \Sigma^k \to \Gamma^n$ where $\Sigma$ and $\Gamma$ are finite sets) with the property that if we are given a string $y$ that is "close" to a valid encoding $C(x)$,[1] then it is possible to find out the message $x$ from the "corrupted encoding" $y$. Towards this goal, it is necessary and sufficient that for any two different messages $x, x'$, their encodings $C(x)$ and $C(x')$ differ in a lot of coordinates. Error-correcting codes are motivated by the task of reliably sending information over noisy channel. In such an application, the sender has a message $x$, he computes and sends $C(x)$ over the channel, because of noise the receiver receives a string $y$ that differs from $C(x)$ in a few coordinates (the ones where transmissions errors occurred), but if the number of errors is bounded then the receiver can still reconstruct $x$ from $y$.

## 1.1 Early Uses of Error-Correcting Codes in Cryptography and Complexity

A natural application of error-correcting codes in computational complexity is to the setting of fault-tolerant computation. In one natural model of fault-tolerant computation, we want to compute a boolean function using a circuit, and each gate of the circuit has a small probability $\epsilon$ of failing (and producing a wrong output). We would like to construct a "fault-tolerant" circuit that, even in the presence of these errors, will have a reasonably high probability of computing the function correctly. (In this model one typically assume that the failures of different gates are mutually independent events.) This problem was introduced by von Neumann [vN56], who suggested that error-correcting codes could be applied to it. Low-density parity-check codes were applied to compute linear functions [Eli58, Tay68] in variants of this model and general functions [Pap85] in the general model.

Another early application of error-correcting codes to cryptography was Shamir's secret sharing scheme [Sha79], which can be seen as an application of Reed-Solomon codes.[2] A different use of coding theory for secret sharing is in [BOGW88] and in subsequent work on the "information-theoretic" model of security for multi-party computations.

Finally, we mention that McEliece's cryptosystem [McE78] is based on the conjectured intractability of certain coding-theoretic problems. The study of the complexity of coding-theoretic problem is clearly an important source of interaction between coding theory and complexity theory, but in this paper we will restrict ourselves to the use of algorithmic coding-theoretic results in complexity theory.

---

[1] meaning that $y$ and $C(x)$ differ in a small number of coordinates.  
[2] This connection was first noticed by McEliece and Sarwate[MS81].

## 1.2 Error-Correcting Codes and Average-Case Complexity

A paper by Levin [Lev87] contains one of the earliest use of coding theory in order to prove an average-case complexity result. The goal of the paper is to construct pseudorandom generators from certain one-way functions and a preliminary step is to construct "hard-core predicates" for such functions.[3] Without getting too technical, we can abstract the use of error-correcting codes in [Lev87] as follows: (i) there is a computational problem $P$ that is presumably hard to solve on a certain set of inputs; (ii) we think of the right answers for $P$ on those inputs as our "message" and we encode it with an error-correcting code; (iii) we define a new computational problem $P'$, which is to compute entries of the above encoding. The important idea is now to observe that if we have a good-on-average algorithm for $P'$, that is, an algorithm that solves $P'$ on all but a small fraction of inputs, we can think of the set of outputs of this algorithm as being a "corrupted" version of our encoding of $P$; using a decoding algorithm for our code we can now solve $P$ correctly on all inputs, which contradicts our intractability assumption for $P$. In conclusion, from a problem $P$ that was assumed to be worst-case hard and from an error-correcting code we have constructed a new problem $P'$ that is average-case hard.

The above outline skips an important point: presumably the complete description of $P$ (and $P'$) is an object of exponential size, while we would like our worst-case to average-case reduction to run in polynomial time, so that we want to use an error-correcting code for which decoding can be performed in *poly-logarithmic* time.

Roughly speaking, in Levin's paper and in other similar cryptographic applications one applies an encoding "locally," to small pieces of the computational problem, so that it is not necessary to have a poly-logarithmic time decoder. As a consequence, the reduction relates a stronger form of average-case complexity to a weaker form (which is enough for these cryptographic applications) instead of relating average-case complexity to worst-case complexity (which is important in other complexity-theoretic applications).

Sections 3 and 4 are devoted to error-correcting codes having decoding algorithms running in poly-logarithmic (or even constant) time, and their applications to complexity theory and cryptography. Some of the applications follow the same line of reasoning sketched above.

## 1.3 Program Testing, Hard-Core Bits, and Sub-linear Time Error-Correction

Work done in the late 1980s and early 1990s on "hiding instances from oracles" [BF90], on the self-reducibility of the permanent [Lip90], and of PSPACE-complete and EXP-complete problems [FF93], as well as work more explicitly focused on average-case complexity [BFNW93] is now seen as based on sub-linear time decoding algorithms for certain polynomial-based error-correcting codes, although this is a view that has become common only since the late 1990s.

Such results were typically discussed in terms of *self-correction*, a notion introduced by Blum, Kannan, Lipton and Rubinfeld [BK89, Lip90, BLR93] in the setting of program testing.

Around the same time, Goldreich and Levin [GL89] introduced an efficient and general way of constructing hard-core predicates for one-way functions (the cryptographic problem mentioned above and extensively discussed in Section 4). The Goldreich-Levin construction is now seen as a sub-linear time list-decoding algorithm for an error-correcting code, a perspective first suggested by Impagliazzo and Sudan in unpublished papers in the early 1990s. The coding-theoretic perspective is useful because it suggests that different, and possibly even more efficient, hard-core predicates can be constructed using different codes and different decoding algorithms.[4] Improvements to [GL89] via the solution of other decoding problems are reported in [GRS00], with an explicit discussion of sub-linear time decoding. Recent work by Akavia, Goldwasser and Safra [AGS03] gives a coding-theoretic interpretation (along with generalizations and improvements) for other hard-core predicate constructions that previously seemed to require ad-hoc algebraic analyses and to be independent of coding theory.

---

[3]We extensively discuss hard-core predicates for one-way permutations in Section 4.

A paper by Babai et al. [BFLS91] is probably the first one to explicitly discuss sub-linear time decoding algorithms for error-correcting codes, and their possible relevance in the classical setting of coding theory, that is, error-resistant storage and transmission of information.[5] The relevance of sub-linear time decoding to average-case complexity, and the generality of the approach of using a code to encode the description of a computational problem, are pointed out explicitly in [STV01]. Katz and this author [KT00] give the first negative results for codes with sub-linear time decoding algorithms and note that, besides their relation to hard-core predicates and average-case complexity, they are also related to private information retrieval [CGKS98], a type of cryptographic protocol discussed in Section 3.2.

## 1.4 Program Testing and Locally Testable Codes

Apart from Levin's work [Lev87], which motivated [GL89], most of the line of work described in the previous section can be traced to the work on program testing by Blum and Kannan [BK89] and Lipton [Lip90]. Suppose that we are interested in computing a function $f$, and that we are given an algorithm $A$ that may or may not be correct: is it possible to test the correctness of $A$ "on the fly" while we run it? The approach proposed in [BK89, Lip90] was roughly as follows: to construct a *self-testing* procedure for $f$ that, given an algorithm $A$, would accept with high probability if $A$ solves $f$ correctly on all inputs, and it would reject with high probability if $A$ is incorrect on many inputs. (Note that the self-testing procedure may accept with high probability an algorithm that makes few mistakes.) An algorithm rejected by the self-tester would be discarded as buggy. If an algorithm $A$ is accepted by the self-tester, then we would use $A$ in connection with a *self-corrector* for $f$. A self-corrector for $f$ is a procedure that given an algorithm $A$ that solves $f$ on many input, and given an arbitrary input $x$, computes $f(x)$ with high probability.

Sudan's PhD Thesis [Sud92] is an early work that makes an explicit connection between self-testing and error-detection [6] and between self-correcting and error-correction.

We note that self-correction, besides being related to error-correction, also relates to average-case complexity (a worst-case intractable problem that is self-correctable is also necessarily average-case intractable). Lipton [Lip90] presents a self-corrector that works for any function that can be expressed as a low-degree polynomial, and, in particular, is a self-corrector for the permanent. Encoding PSPACE-complete and EXP-complete problems using a polynomial-based encoding (which is called the Reed-Muller code, as we will see in a later section), Feigenbaum and Fortnow [FF93] give self-correctors for certain PSPACE-complete and EXP-complete problems, and Babai et al [BFNW93] use these results to prove average-case complexity results for certain EXP-complete problems. Since the self-correction perspective is very natural, it took some time to see the constructions of [FF93, BFNW93] as being about error-correcting codes with sub-linear time decoding.

---

[4]Such improvements were the focus of the manuscripts by Impagliazzo and Sudan.

[5]As we discuss below, known and conjectured negative results make such applications unlikely.

[6]The *error-detection* problem for an error correcting code is to distinguish a valid encoding $C(x)$ from a string $y$ that is not a valid encoding of any message.

Just like self-correcting is strongly related to sub-linear time decoding of error-correcting codes, so self-testing is related to sub-linear time error-detection. The self-testing algorithms by Blum, Luby and Rubinfeld [BLR93] for linear functions and by Gemmel et al. [GLR+91, RS96] for polynomial functions can indeed be see as sub-linear time error-detection algorithms for certain error-correcting codes. Such testing algorithms played a fundamental role in the construction of probabilistically checkable proofs (PCP) [FGL+91, AS98, ALM+98], which in turn revolutionized the study of approximation algorithms.[7]

Locally testable codes, that is, codes with sub-linear time error-detection algorithms, were soon recognized to be the combinatorial core of PCP construction, and the question of providing simpler and more efficient constructions of such codes was posed as an open question in various writings from the mid 1990s, such as [Aro94, Spi95, FS95, RS96]. Great progress has been made towards such constructions in the past two years, with the latest results [BSGH+04, DR04] providing a more clarifying perspective on the relationship between these codes and PCP constructions.

## 1.5 Further Reading

Regarding coding theory in general, van Lint's book [vL99] is an excellent reference. Madhu Sudan's notes [Sud, Sud01] are excellent introductions to algorithmic coding theory, and they are the main source that we used for our brief presentation of results in algorithmic coding theory in Section 2.

A survey on applications of coding theory to complexity theory was written by Joan Feigembaum [Fei95] about ten years ago. Many themes treated in [Fei95] are still current. Venkat Guruswami's thesis [Gur01] has a chapter on applications of coding theory to complexity and cryptography. A survey paper by Madhu Sudan [Sud00] focuses on applications of list-decoding algorithms to complexity theory, including the applications to average-case complexity and hard-core predicates that we discuss in this paper.

## 1.6 Organization of this Paper

We start the paper with some review material on error-correcting codes and algorithmic coding theory. This material has wider applications than the ones that we chose to focus on in this paper.

We then consider sub-linear time error-correction algorithms, their relation to private information retrieval, and their applications in average-case complexity and cryptography.

Finally we discuss sub-linear time error-detection algorithms and their relation to PCP constructions.

## 2 Error-Correcting Codes

### 2.1 Shannon's Setting

A party called the *sender* has a message $x \in \Sigma^k$ that he wants to send to another party called the *receiver*. Here $\Sigma$ is a finite alphabet (often $\Sigma = \{0,1\}$) and $k$ is the message length.

The sender and the receiver communicate through a *noisy channel* that introduces errors. To eliminate errors (or, at least, to dramaticallt reduce the probability of errors) the sender first encodes his message using an encoding function $C : \Sigma^k \to \Sigma^n$ with $n > k$ that introduces some redundancy, and then sends $C(x)$ through the channel. The receiver receives a string $y$ that is possibly different from $C(x)$ because of transmission errors. The receives then feeds $y$ to a *decoding algorithm* $D$ that, under some assumption about the error pattern introduced by the channel, is able to compute $x$.

---

[7]This is a story that is both too long and too exciting to be effectively summarized here. We try to give such a summary in Section 5.4. The reader should also refer to one of the many excellent survey papers on the subject, such as, for example, [Aro98].

We would like to design efficient procedures $C$ and $D$ such that the above holds under general assumptions about the channel and with $n$ not much larger than $k$. This setting was introduced by Shannon [Sha48] in his monumental work that defined information theory.

## 2.2 Error-Correcting Codes

The *Hamming Distance* $d_H(a,b)$ between two strings $a,b \in \Sigma^n$ is the number of entries $i$ such that $a_i \neq b_i$.

An $[n,k,d]_q$ code is a function $C : \Sigma^n \to \Sigma^k$ such that

- $|\Sigma| = q$;
- For every $x, x' \in \Sigma^k$, $d_H(C(x), C(x')) \geq d$.

The paramerer $k$ is called the *information length* of the code and $n$ is called the *block length*. Abusing terminology a little, we call $d$ the *minimum distance* of the code.[8]

If a $[n,k,d]_q$ code admits a decoding procedure that is always able to correct $e$ errors, then it must be $d \geq 2e + 1$. Conversely, if $d \geq 2e + 1$ then there is a (possibly not efficiently computable) decoding procedure that is able to correct up to $e$ errors.

Error-correcting codes, introduced by Hamming [Ham50], solve the coding problem in models where there is a upper bound to the number of errors introduced by the channel. Error-correcting codes can also be used in settings where we have a probabilistic model for the channel, provided that we can show that with high probability the number of errors introduced by the channel is smaller than the number of errors that the decoding procedure can correct. In the rest of this paper we only discuss error-correcting codes, but the reader can see that any algorithmic result about error-correcting codes implies an algorithmic solution to Shannon's problem for various distributions of errors.

For a given $k$, we are interested in constructing $[n,k,d]_q$ codes where $n$ is small (ideally, $n = O(k)$), $d$ is large (ideally, we would like $d = \Omega(n)$) and $q$ is small (ideally, $\Sigma = \{0,1\}$ and $q = 2$). Sometime we will call the ratio $k/n$ the *information rate* (or just *rate*) of the code, which is the "amortized" number of alphabet elements of the message carried by each alphabet element sent over the channel. We will also call the ratio $d/n$ the *relative minimum distance* of the code.

## 2.3 Negative Results

Before seeing constructions of error-correcting codes, let us start by seeing what kind of trade-offs are impossible between $k$, $d$ and $n$.

Suppose $C$ is a $[n,k,d]_q$ code, and associate to each message $x$ the set of strings $S_x$ defined as the set of all strings $y$ that agree with $C(x)$ on the first $n - d + 1$ coordinates. We claim that these sets are all disjoint. Otherwise, if we had $y \in S_x \cap S_{x'}$ we would have that $y$ and $C(x)$ agree in the first $n-d+1$ coordinates, and so would $y$ and $C(x')$, but then $C(x)$ and $C(x')$ would also have to agree on the firt $n-d+1$ coordinates and this would contradict the minimum distance requirement of the code. Now, we have $q^k$ disjoint sets each of size $q^{n-d+1}$ contained in a space of size $q^n$, and so we have proved the following result.

---

[8]More precisely, the minimum distance of a code $C$ is $\min_{x \neq x'}\{d_H(C(x), C(x'))\}$, so that if $C$ is an $[n,k,d]_q$ code, then $d$ is a *lower bound* to the minimum distance.

**Lemma 1 (Singleton Bound)** *In a $[n, k, d]_q$ code, $k \leq n - d + 1$.*

As we will see later, this negative result can be matched if the size $q$ of the alphabet is large enough compared to $n$. For smaller alphabets, however, stronger bounds are known.

**Lemma 2 (Plotkin's Bound)** *In a $[n, k, d]_q$ code, $k \leq n - (q/(q-1))d + log_q n$.*

For example, if $q = 2$, then the relative minimum distance $d/n$ cannot be larger than $1/2$, and for constant $q$ it cannot be larger than $1 - 1/q$. For proofs of these results, see for example [vL99].

## 2.4 Constructions of Error-Correcting Codes and Decoding Algorithms

In this section we will consider various constructions of error-correcting codes. All these constructions will have the property of being *linear*, that is the alphabet $\Sigma$ will be a field $\mathbb{F}$, and the encoding function $C : \mathbb{F}^k \to \mathbb{F}^n$ will be a linear function.

If $C$ is a linear code, then there is a matrix $A$ such that the encoding function can be specified as $C(x) = A \cdot x$. Also, there is a matrix $H$ such that $y$ is a codeword (that is, a possible output of $C$) if and only if $H \cdot y = \mathbf{0}$, where $\mathbf{0}$ is the all-zero vector. This means that for every linear code $C$ there is always an encoding circuit of size at most quadratic (that simply computes $A \cdot x$ given $x$) and a circuit of size at most quadratic that solves the *error-detection* problem, that is, the problem of deciding whether a given string is a codeword or not.

Let the *weight* of a vector $y \in \mathbb{F}^n$ be defined as the number of non-zero entries. (Equivalently, the weight of a vector is its Hamming distance from the all-zero vector.) Then it is easy to see that the minimum distance of a linear code is equal to the minimal weight of a non-zero codeword.[9] This observation often simplifies the study of the minimum distance of linear codes.

### 2.4.1 Random Error-Correcting Codes

As a first example of linear error-correcting code, we see what happens if we pick at random a linear code over the field $\{0, 1\}$. In order to show that, with high probability, the code has large minimum distance, we show that, with high probability, all non-zero inputs are mapped into codewords with a large number of ones. This is easy to show because, for a random matrix $A$ and a fixed non-zero vector $x$, the encoding $A \cdot x$ is uniformly distributed, and so it has a very low probability of having low weight. The argument is completed by using a union bound. The formal statement of the result and a sketch of the proof is below. This existence result is called the Gilbert-Varshamov bound because it was first proved by Gilbert [Gil52] for general random codes, and then Varshamov [Var57] observed that the same bond could be obtained by restricting oneself to random linear codes.

**Lemma 3 (Varshamov)** *For every $\delta < 1/2$ and every $n$ there is a $[n, Rn, \delta n]_2$ linear code such that*
$$R \geq 1 - H_2(\delta)) - \Theta((\log n)/n)$$

---

[9]To be precise, one also needs to assume that the encoding function $C$ is injective. Alternatively, one can see that the minimum distance is equal to the minimum weight of the encoding of a non-zero input.

Where $H_2(x) = x\log_2(1/x) + (1-x)\log_2(1/(1-x))$ is the binary entropy function.

PROOF: We pick a linear function $C : \{0,1\}^k \to \{0,1\}^n$ at random by picking at random a $k \times n$ 0/1 matrix $A$ and defining $C(x) = Ax$.

We use the probabilistic method to show that there is a positive probability that every non-zero message is encoded into a string with at least $d$ ones.

For a particular $x \neq 0^k$, we note that $C(x)$ is uniformly distributed in $\{0,1\}^n$, and so

$$\mathbf{Pr}[w(C(x)) < d] = 2^{-n} \cdot \sum_{i=0}^{d-1} \binom{n}{i} \leq 2^{-n} \cdot 2^{nH_2(d/n)+O(\log n)}$$

Where we used the fact that there are $2^{n \cdot H(k/n)+\Theta(\log n)}$ binary strings of length $n$ having weight $k$.

A union bound shows that

$$\mathbf{Pr}[\exists x \neq 0^k : w(C(x)) < d] < 2^k \cdot 2^n \cdot 2^{nH_2(d/n)+O(\log n)}$$

which is smaller than 1 under the assumption of the lemma. □

It is worth noting that the observation about minimum distance versus minimum weight plays a role in the proof. A proof that didn't use such a property would have considered the event that two possible inputs $x, x'$ are mapped into encodings of distance smaller than $d$, and then we would have taken a union bound over all such pairs. This would have let to a bound worse by a factor of two than the bound we achieved above.

As an interesting special case, we have that for every $\epsilon > 0$ there is a constant rate $R = O(\epsilon^2)$ such that for every $n$ there is a $[n, Rn, (1/2 - \epsilon) \cdot n]_2$ linear code. That is, there are linear codes with constant rate and with relative minimum distance arbitrarily close to $1/2$. (Recall that Plotkin's bounds does not permit codes with relative minimum distance strictly larger than $1/2$.) More generally, if $q$ is a prime power and $\epsilon > 0$ then there is a constant rate $R$ for which linear $[n, Rn, (1 - 1/q - \epsilon) \cdot n]_q$ linear codes exist for all sufficiently large $n$.

There is no known algorithm to decode random linear codes in polynomial time on average. It is, however, possible to solve the decoding problem for any linear code in exponential time by using brute force.

### 2.4.2 Reed-Solomon Codes

The next code we consider is based on the following well-known fact about (univariate) polynomials: a polynomial of degree $t$ is either identically zero or it has $\leq t$ roots.

**Encoding and Minimum Distance.** In a Reed-Solomon code [RS60] we think of every message as representing a low-degree polynomial, and the encoding of the message is the $n$ values that we get by evaluating of the polynomial at $n$ fixed points. A more formal description follows.

Let $q$ be a prime power and $\mathbb{F}_q$ be a finite field of size $q$. Let us fix $n$ distinct elements of $\mathbb{F}_q$, $x_1, \ldots, x_n$, and let $k < n$. We define a $[n, k, n-k+1]_q$ linear code as follows.

Given a message $(c_0, \ldots, c_{k-1})$, we interpret it as a description of the polynomial $p(x) = c_0 + c_1 x + \ldots + c_{k-1} x^{k-1}$. The encoding of such a message will be the vector $(p(x_1), \ldots, p(x_n))$.

Such a procedure maps indeed a message of length $k$ into an encoding of length $n$, and it is a linear mapping. To verify the claim about the minimum distance, if $(c_0, \ldots, c_{k-1})$ is not the all-zero vector, then the corresponding polynomial $p$ is a non-zero polynomial of degree $k-1$. Such a polynomial can have at most $k-1$ roots, and so at lest $n-(k-1)$ of the values $p(x_1), \ldots, p(x_n)$ must be non-zero. The reader should know that Reed-Solomon codes meet the Singleton bound, and thus have an optimal trade-off between rate and minimum distance.

**Decoding Algorithms.** Decoding the Reed-Solomon code in a channel that introduces $e < (n-k+1)/2$ errors is equivalent to the following problem:

- Given: distinct elements $x_1, \ldots, x_n$ of $\mathbb{F}_q$, parameters $e$ and $k$, with $e < (n-k+1)/2$, and elements $y_1, \ldots, y_n$ of $\mathbb{F}_q$;

- Find: a polynomial $p$ of degree at most $k-1$ such that

$$\#i : p(x_i) \neq y_i \leq e$$

Note that, because of the constraint on $e$ and $k$, the problem has always a unique solution $p$. A polynomial time algorithm for the decoding problem has been known since the early 1960s, following Peterson's polynomial time algorithm to decode BCH codes [Pet60] and the reduction of Gorenstein and Zierler [GZ61], who showed that decoding Reed-Solomon codes can be seen as a special case of the problem of decoding BCH codes.[10] A simple and efficient polynomial time algorithm for the decoding problem for Reed-Solomon codes was devised by Berlekamp and Welch [WB86]. We describe the Berlekamp-Welch algorithm in the Appendix.

### 2.4.3 Reed-Muller Codes

Reed-Muller codes [Ree54] generalize Reed-Solomon codes by considering *multivariate* polynomials instead of univariate polynomials. That is, we think of each message as specifying a low-degree multivariate polynomial, and the encoding of the message is the evaluation of the polynomial at a certain set of points. If the evaluation points are suitably chosen, we still have the property that a non-zero low-degree polynomial has few roots among these points, and so we can still infer that the resulting encoding is an error-correcting code with good minimum distance.

Let $q$ be a prime power and $\mathbb{F}$ be a field of size $q$. To define a Reed-Muller code we choose a subset $S \subseteq \mathbb{F}$, a degree $t < |S|$ and a parameter $m$. We will think of an input message as the description of an $m$-variate degree-$t$ polynomial. The message is encoding by specifying the value of the polynomial at all the points in $S^m$.

We can see that there are up to $\binom{m+t}{m}$ possible monomials in an $m$-variate polynomial of degree at most $t$. An input message is, therefore, a sequence of $k = \binom{m+t}{m}$ coefficients. The encoding is the evaluation of the polynomial at $n = |S|^m$ different points. Note that if $m = 1$ we are back to the case of Reed-Solomon codes. Regarding minimum distance, we have the following result, that is called the Schwartz-Zippel Lemma (after [Sch80] and [Zip79]) in the computer science literature.

**Lemma 4** *If $p$ is a non-zero degree-$t$ polynomial over a field $\mathbb{F}$ and $S \subseteq \mathbb{F}$, then*

$$\mathbf{Pr}_{x \sim S^m}[p(x) = 0] \leq \frac{t}{|S|}$$

---

[10] BCH codes are a class of algebraic error-correcting codes that we will not discuss further in this paper.

To compare Reed-Solomon codes and Reed-Muller codes it can be helpful to look at a concrete example. Suppose we want to map $k$ field elements into $n$ field elements and we want the minimum distance to be at least $n/2$.

In the Reed-Solomon code we would choose $n$ to be $2k$, and so the minimum distance will be $k+1 > n/2$. The field size has to be at least $2k$.

In the Reed-Muller code, we also have to choose the parameter $m$. Suppose we choose $m = 2$. Then we want to choose $t$ and $S$ such that $t = |S|/2$, $k = \binom{t+2}{2}$, so that $k \approx t^2/2$ and $|S| = 2t$. The encoding length is $|S|^2 = 4t^2 \approx 8k$. The field size has to be at least $2t \approx 2\sqrt{2k}$.

We see that the rate has become worse, that is, the encoding length is bigger, but the field size can be smaller, that is, a smaller alphabet is sufficient.

For larger values of $m$, we would get an encoding length $n = 2^{O(m)}k$ and a requirement that the field be of size at least $2^{O(m)} \cdot k^{1/O(m)}$.

What is the extreme case of very small alphabet and very large encoding length? We can choose $t = 1$, so that we only need $|S| = 2$, but then we have $m = k - 1$, and, catastrophically, $n = 2^{k-1}$. In this code, we see the input message $(c_0, c_1, \ldots, c_{k-1})$ as representing the affine function $(x_1, \ldots, x_{k-1}) \to c_0 + c_1 x_1 + \cdots + c_{k-1} x_{k-1}$. The encoding is the evaluation of such a function at all points in $\{0,1\}^{k-1}$.

We may in fact consider an even more wasteful code in which we interpret the message as a linear, instead of affine, function. That is, we think of a message $(c_1, \ldots, c_k)$ as representing the function $(x_1, \ldots, x_k) \to c_1 x_1 + \cdots + c_k x_k$, and the encoding is the evaluation of such a function at all points in $\{0,1\}^{k-1}$. Such a $[k, 2^k, 2^{k-1}]_2$ encoding is typically called (with some abuse of terminology) the *Hadamard code*.

### 2.4.4 Concatenated Codes

Reed-Solomon and Reed-Muller codes have very good trade-offs between rate and minimum distance and, indeed, the Reed-Solomon codes exhibit an *optimal* trade-off. The drawback of Reed-Solomon and Reed-Muller codes is the need for large alphabets: in the Reed-Solomon code the alphabet size must be at least as large as the encoding length; in the Reed-Muller codes smaller alphabets are possible, but the trade-off between rate and minimum distance worsens when one uses smaller alphabets.

*Concatenation* is a method that can be used to reduce the alphabet size without compromising too much the information rate and the minimum distance.

Suppose that we have a $[N, K, D,]_Q$ code $C_o : \Gamma^K \to \Gamma^N$ and a $[n, k, d]_q$ code $C_i : \Sigma^k \to \Sigma^n$. Suppose also that $Q = q^k$, and let us fix some way to identify elements of $\Gamma$ with strings in $\Sigma^k$. We call $C_o$ the *outer* code and $C_i$ the *inner* code.[11]

Let $X \in \Gamma^k$ be a message and let $C_o(X)$ be its encoding. We can think of each coordinate of $C_o(X)$ as containing a message from $\Sigma^k$, and we can apply the encoding $C_i()$ to each such message. The end result will be a string in $\Sigma^{Nn}$. If we start from two different messages $X, X'$, their encodings $C_i(X)$ and $C_i(X')$ will differ in at least $D$ coordinates, and each such coordinate will lead at least $d$ coordinates of its second-level encoding to be different. In summary, we have described a way to encode a message from $\Gamma^K$ as a string in $\Sigma^{nN}$ so that any two different encodings differ in at least $dD$ coordinates. If we observe that we can identify $\Gamma^K$ wih $\Sigma^{kK}$, we conclude that what we just described is a $[nN, kK, dD]_q$ code.

---
[11]The reason for this terminology will be clear in a minute.

**Lemma 5 (Concatenation)** *Suppose we have an explicit construction of a $[N, K, D,]_Q$ code and of a $[n, k, d]_q$, with $Q = q^k$, then we also have an explicit construction of a $[nN, kK, dD]_q$ code.*

This idea is due to Forney [For66].

By concatenating a Reed-Solomon code of rate $1/2$ and relative minimum distance $1/2$ with another Reed-Solomon code with the same rate and relative minimum distance, we can get, say, a $[n, n/4, n/4]_{O(\log n)}$ code.

If we concatenate such a code with a linear code promised by the Gilbert-Varshamov bound, we get a $[n, \Omega(n), \Omega(n)]_2$ code, and the needed binary code is so small that it can be found efficiently by brute force.

What about decoding? It is easy to see that if we concatenate a $[N, K, D,]_Q$ code and a $[n, k, d]_q$ code, and if the outer code (respectively, the inner code) has a decoder algorithm that can correct $E$ errors (respectively, $e$ errors), then it is easy to design a decoding algorithm for the concatenated code that corrects up to $eE$ errors.

Unfortunately this is far from optimal: since $e < d/2$ and $E < D/2$, we are able to correct $< dD/4$ errors, while we might hope to correct up to $dD/2 - 1$ errors.

There is a more sophisticated general decoding algorithm for concatenated codes, due to Forney [For66], which is beyond the scope of this short overview. Forney's algorithm can indeed decode up to $dD/2 - 1$ errors.

### 2.4.5 Error Rate for Which Unique Decoding is Possible

Using concatenation, the codes described so far, algorithms for these codes, and Forney's decoding algorithm for concatenated codes, it is possible to show that for every $\epsilon > 0$ there is a rate $R$ and, for every large enough $n$, a $[n, Rn, (1/2 - \epsilon) \cdot n]_2$ code that can be encoded and decoded in polynomial time; the decoding algorithm is able to correct up to $(1/4 - \epsilon/2) \cdot n$ errors.

This is the largest fraction of errors for which the decoding problem can be solved. Recall that the decoding problem is well defined only if the number of errors is less than half the minimum distance, and that for a binary code with good rate the minimum distance cannot be more than $n/2$, so that it is not possible to correct more than $n/4$ errors in a binary code.

If we use an alphabet of size $q$, results described in this section lead to the result that for every $\epsilon > 0$ there is a rate $R$ and, for every large enough $n$, a $[n, Rn, (1 - 1/q - \epsilon) \cdot n]_q$ code that can be encoded and decoded in polynomial time; the decoding algorithm is able to correct up to $(1/2 - 1/2q - \epsilon/2) \cdot n$ errors. This is, again, essentially the best possible fraction of errors for which unique decoding is possible

## 2.5 List Decoding

The notion of *list decoding*, first studied by Elias [Eli57], allows us to break the barrier of $n/4$ errors for binary codes and $n/2$ errors for general code.

If $C : \Sigma^k \to \Sigma^n$ is a code, a list-decoding algorithm for radius $r$ is an algorithm that given a string $y \in \Sigma^n$ finds all the possible messages $x \in \Sigma^k$ such that the Hamming distance between $C(x)$ and $y$ is at most $r$. If $r$ is less than half the minimum distance of the code, then the algorithm will return either an empty list or the unique decoding of $y$. For very large values of $r$, the list of possible decodings could be exponentially big in $k$. The interesting combinatorial question, here, is to find codes such that, even for very large values of $r$, the list is guaranteed to be small.[12]

Algorithmically, we are interested in producing such lists in polynomial time.

As we will see, there are binary codes for which efficient list-decoding is possible (with lists of constant size) even if the number of errors is of the form $(1/2 - \epsilon) \cdot n$. For codes over larger alphabets, even $(1 - \epsilon) \cdot n$ errors can be tolerated.

### 2.5.1 The Hadamard Code

The simplest case to analyse is the Hadamard code.

**Lemma 6** *Let $f : \{0,1\}^k \to \{0,1\}$ be a function and $0 < \epsilon < 1/2$. Then there are at most $1/4\epsilon^2$ linear functions $l$ such that $f()$ and $l()$ agree in at least a $1/2 + \epsilon$ fraction of inputs.*

This means that good list-decoding is possible, at least combinatorially. From the algorithmic point of view, we can consider the fact that the input for a decoding algorithm is a string of length $n = 2^k$, and there are only $2^k$ possible decoding. Therefore, a brute-force decoding algorithm runs in polynomial time.

In Section 4 we will see a probabilistic algorithm that runs in time polynomial in $k$ and $1/\epsilon$.

### 2.5.2 Reed-Solomon Codes

The list-decoding problem for Reed-Solomon codes can be stated as follows: given $n$ distinct points $(x_1, y_1), (x_2, y_2), \ldots, (x_n, y_n)$ in $\mathbb{F}_q^2$ and parameters $k, t$, find a list of all polynomials $p$ such that:

1. $p$ has degree $\leq k - 1$; and

2. $\# : p(x_i) = y_i \geq t$

With no further constraints on $n, k$ and $t$, it is not clear that the list of such polynomials is small (that is, $\text{poly}(n, k, t)$). In particular, if $t = k$, there are at least $q^k$ such distinct polynomials (pick any of the $k$ points and interpolate). Therefore, we will definitely require that $t > k$ if we would like to efficiently list-decode.

The first polynomial time algorithm for this problem, for $t > \sqrt{2nk}$ is due to Sudan [Sud97]. The error bound was then improved to $t > \sqrt{nk}$ in [GS99], which is tight.[13]

We give a proof of the following theorem in the Appendix.

**Theorem 7 ([Sud97])** *Given a list of $n$ points $(x_1, y_1), \ldots, (x_n, y_n)$ in $\mathbb{F}_q^2$, we can efficiently find a list of all polynomials $p(x)$ of degree at most $k - 1$ that pass through at least $t$ of these $n$ points, as long as $t > 2\sqrt{nk}$. Furthermore, the list has size at most $\sqrt{n/k}$.*

---

[12]We will often work in settings where the size of the list is upper bounded by a constant. A size polynomial in $k$ is also acceptable.

[13]Meaning that for smaller values of $t$ the size of the list may be superpolynomial, and so the problem becomes intractable even from a combinatorial perspective.

### 2.5.3 Concatenated Codes

Suppose we have an outer code $C_o$ which is a $[N, K, D]_Q$ code and an inner code $C_i$ which is a $[n, k, d]_q$ code with $q^k = Q$, and that we define the concatenated code $C$ which is then a $[nN, kK, dD]_q$ code. Suppose we have a good list-decoding algorithm for both the outer code and the inner code: can we derive a list-decoding algorithm for the concatenated code?

Here is a very simple idea: apply the inner list-decoding algorithm to each block, and so come up with a sequence of $N$ lists. Pick a random element from each list, and construct, in this way a string of length $N$, then apply the outer list-decoding algorithm to this list.

Suppose that the inner decoding algorithm was able to decode from $(1-\epsilon)n$ errors and produce a list of size $l$.

Suppose also that, overall, we are given a string that agrees with a valid codeword $C(x)$ of $C$ in at least $2\epsilon nN$ entries. Then there are at least $\epsilon N$ blocks in which there are at most $(1-\epsilon)n$ errors, and in which the inner decoding algorithm has a decoding consistent with $C(x)$ in the list. On average, when we pick randomly from the lists, we create a string that has agreement at least $\epsilon N/l$ with the outer encoding of $x$. If the outer list-decoding algorithm is able to tolerate $(1-\epsilon/l)N$ errors, then we will find $x$ in the list generated by the outer list-decoding algorithm.

This argument can be derandomized by observing that we do not need to choose independently from each list. We deduce that

**Theorem 8** *If $C_o$ is a $[N, K, D]_Q$ code with a $(L, (1-\epsilon/l)N)$ list decoding algorithm, and $C_i$ is a $[n, k, d]_q$ code with a $(l, (1-\epsilon)n)$ list decoding algorithm, and $q^k = Q$, then the concatenated code $C$ is a $[nN, kK, dD]_q$ code and it has a $(L, (1-2\epsilon)nN)$ list decoding algorithm.*

Basically, if both the outer and the inner code can be list-decoded from an arbitrarily large fraction of errors, then so is their concatenation.

Similarly, one can argue that if the outer code can be list-decoded from an arbitrarily large fraction of errors, and the inner code can be list-decoded from a fraction of errors arbitrarily close to $1/2$, then their concatenation can be list-decoded from a fraction of errors arbitrarily close to $1/2$.

More sophisticated algorithms for list-decoding concatenated codes are in [GS00b].

### 2.5.4 Error Rate for Which List-Decoding is Possible

Using the results that we mentioned above we can prove that for every $\epsilon$ and for every $k$ there is a polynomial time encodable code $C : \{0,1\}^k \to \{0,1\}^n$ that is $(L, (1/2 - \epsilon) \cdot n)$ list decodable in polynomial time, where $n = \text{poly}(k, 1/\epsilon)$ and $L = \text{poly}(1/\epsilon)$. Thus, a meaningful (and useful) form of error-correction is possible with a binary code even if the number of errors is close to $n/2$.

By the way, considerably better results are possible, and, in particular, it is possible to have $n = O(k \cdot \text{poly}(1/\epsilon))$, so that the rate is constant for fixed $\epsilon$. It is also possible to implement the list-decoder in nearly-linear or even linear time. (For some recent results in these directions, see e.g. [GS00b, GI01, GI02, GI03] and the references therein.)

For larger alphabets, it is similarly possible to have list-decodable codes that tolerate a fraction of errors close to $(1 - 1/q) \cdot n$.

# 3 Sublinear Time Unique Decoding

In this section we discuss error-correction algorithms that run in sub-linear time, and their relations to private information retrieval (a type of cryptographic protocol) and to average-case complexity, as well as to the notions of self-correction, instance-hiding and random-self-reduction.

## 3.1 Locally Decodable Codes

Let $C : \Sigma^k \to \Sigma^n$ be an error correcting code. The results and the algorithms described so far deal with the following setting: for some message $x \in \Sigma^k$, the codeword $C(x) \in \Sigma^n$ has been "sent," however a corrupted string $y \in \Sigma^n$ has been "received," which differs from $C(x)$ in a bounded number of entries; our goal is to reconstruct $x$, possibly in time polynomial or even linear in $n$. In this section we deal with algorithms whose running time is *sublinear* in $n$ and $k$, or even a *constant* independent of $n$. An algorithm with a such a fast running time cannot possibly reconstruct the entire message, since it does even have to time to write it down. Instead, we will look for algorithms that given an index $i$ and a corrupted version of $C(x)$ will be able to compute just the entry $x_i$. Such codes are called *locally decodable* error-correcting codes.

Such codes could be useful in the setting of information storage: a very large amount of information (for example several songs) could be encoded as a single codeword and then stored in a medium that is subject to become partially corrupted over time (for example a CD, which is subject to scratches). When a particular piece of information (for example, a song) is needed, then the decoding algorithm will not decode the entire content of the medium, but only the part that is needed. Hopefully, then, the decoding time will be proportional only to the length of the desired fragment of information, whereas the whole medium will be robust against a number of errors proportional to the size of the entire storage.[14]

As we will see, however, even the best known locally decodable codes have very poor rate, and this is conjectured to be an inherent problem, and none of them seem to have applications to data transmissions and data storage. In complexity theory and in cryptography, on the other hand, they have several applications, as we will see.

We start with a formal definition. We make the following convention, that we maintain through the paper: whenever we refer to an oracle algorithm, we assume that the algorithm makes *non-adaptive* queries.

**Definition 1 (Locally Decodable Code)** *A code $C : \Sigma^k \to \Gamma^n$ is $(q, \delta, p)$-locally decodable if there is a probabilistic oracle algorithm of query complexity at most $q$ such that for every message $x \in \Sigma^k$, index $i \in \{1, \ldots, k\}$, and string $y$ such that $d(y, C(x)) \leq \delta n$ we have*

$$\mathbf{Pr}[A^y(i) = x_i] \geq p \ .$$

*The probability is taken over the internal coin tosses of $A$.*

---

[14] In practice, the information on music and data CD and on DVD is encoded in a different way. The data is split in relatively small blocks, each block is encoded with a variant of the Reed-Solomon code, and then the encoding of each block is scattered in non-consecutive locations on the disk. This system has very poor resistance against worst-case errors, because one can destroy a block of the original data by damaging its encoding, which is a very small fraction of the overall encoding. On the other hand, this system performs well against a small number of "burst" errors, in which contiguous locations are damaged. The latter type of errors is a good model for the damage suffered by CDs and DVDs

In this setting, a message $x$ made of $k$ elements of an alphabet $\Sigma$ is encoded as $n$ elements of alphabet $\Gamma$. After at most $\delta n$ errors occur, we are interested in reconstructing an entry $x_i$ of $x$. Algorithm $A$ performs such a task with probability at least $p$ while looking at only $q$ entries of the corrupted encoding of $x$.

We will mostly be interested in the case in which $q$ is a small constant, $\Sigma = \{0,1\}$, and $\Gamma = \{0,1\}^t$.

## 3.2 Relation to Private Information Retrieval

A *private information retrieval scheme* is a system in which a "database" $x$, a $k$-bit string, is known to $q$ independent "servers" $S_1, \ldots, S_q$. A "user" is interested in a bit $x_i$ of $x$ and wants to retrieve it with a single round of communication, in such a manner that no server can tell the value $i$. More formally, we require that the distribution of the query sent to each server be independent of $i$. A weaker requirement is that the distributions corresponding to various $i$ be statistically close in the view of each server.

**Definition 2 (One-Round Private Information Retrieval)** *A $(1-\delta)$-secure $q$-server one-round private information retrieval system with recovery probability $p$ for $k$-bits database is a collection of $q+2$ procedures $(Q, S_1, \ldots, S_q, R)$ that work as follows.*

*Fix a string $x \in \{0,1\}^k$ and an index $i \in [k]$. On input an index $i$ and random coins, the query procedure $Q$ computes $q$ queries $j_1, \ldots, j_q \in [n]$. On input the query $j_t$ and the string $x \in \{0,1\}^k$, the $t$-th server produces (deterministically) the answer $a_t = S_t(x, j_t) \in \{0,1\}^l$. Given $i$, the recovery procedure $R$ computes $R(i, a_1, \ldots, a_q)$.*

*We require that the following two conditions hold:*

**Recovery** *For every $x \in \{0,1\}^k$ and $i \in [k]$, there is a probability at least $p$ (over the coin tosses of $Q$) that the final output of $R$ equals $x_i$.*

**Privacy** *For every $x \in \{0,1\}^k$, every $i, j \in [k]$ and every $t \in [q]$, if we sample at random $(a_1, \ldots, a_q) \sim Q(i)$ and $(a'_1, \ldots, a'_q) \sim Q(j)$, then the distribution of $a_t$ and $a'_t$ have statistical distance at most $\delta$.*

We call $l$ the *answer size* of the PIR system, and $\log_2 n$ the *query length*. The communication complexity of the system is $q \cdot (l + \log n)$.

The definition is long and technical, but hopefully it is not too hard to follow it if one has clearly in mind the intuitive notion that the definition is trying to capture. All known constructions have perfect recovery probability $p = 1$ and they are 1-secure (that is, queries made to the same server for two different indices are identically distributed).

A superficial connection between PIR and LDCs of constant query complexity is that all known constructions of both object follow from constructions of a more general object, that we define below.

**Definition 3 (Perfectly Smooth Decoder)** *A perfectly smooth decoder for a code $C : \Sigma^k \to \Gamma^n$ is a probabilistic oracle algorithm $A$ such that for every $i \in k$ and every $x \in \{0,1\}^k$ we have*

$$\mathbf{Pr}[A^{C(x)}(i) = x_i] = 1$$

*Furthermore, if $q$ is the query complexity of $A$, then for every $j \in [q]$ and every $i \in [k]$, the distribution of the $j$-th oracle query made by $A^{C(x)}(i)$ is uniform over $[n]$.*

| Setting | Construction of perfectly smooth codes | Lower Bounds for all LDCs |
|---|---|---|
| 2 queries, Boolean encoding | $n = 2^k$ | $n = 2^{\Omega(k)}$ [KdW03] |
| 2 queries, encoding using $\{0,1\}^l$ | $n = l \cdot 2^{k/l}$ | $n = 2^{\Omega(k/2^{\text{poly}\,l})}$ [KdW03] |
| 2 queries, encoding using $\{0,1\}^{O(k^{1/3})}$ | $n = 2^{O(k^{1/3})}$ [CGKS98] | $n = \Omega(k^{4/3})$ [KT00] |
| 3 queries, Boolean encoding | $n = 2^{\sqrt{k}}$ [BI01] | $k = \tilde{\Omega}(n^2)$ [KdW03] |
| 3 queries, encoding using $\{0,1\}^l$ | $n = 2^{\sqrt{k/l}}$ [BI01] | $k = \Omega((n/l)^{1.5})$ [KT00] |
| $q$ queries, Boolean encoding | $n = 2^{k^{O(\log \log q / q \log q)}}$ [BIKR02] | $k = \tilde{\Omega}(n^{q/(q-2)})$ [KdW03] |

Table 1: Main known results on Locally Decodable Codes with decoders of constant query complexity.

Suppose we have a perfectly smooth decoder for a code $C$. Then it is easy to see that for every $\delta < 1/q$ the decoder shows that $C$ is also a $(q, \delta, 1 - \delta q)$ locally decodable code. If $y$ is a string that is $\delta$-close to a codeword $C(x)$, and let $i$ be any index; then there is at least a $1 - \delta q$ fraction of the coin tosses of $A$ such that the view, and outcome, of $A^y(i)$ and $A^{C(x)}(i)$ are the same, and so $A^y(i)$ has at least a probability $1 - \delta q$ of correctly computing $x_i$. If $q$ is not a constant, then this observation does not give us a LDC that is able to correct a constant fraction of errors, so only perfectly smooth decoder of constant query complexity give good LDCs.

Regarding PIR, consider the following approach. The user simulates $A^{(\cdot)}(i)$ and finds out the queries $j_1, \ldots, j_q$ that $A$ would have made.[15] Then it sends each query to a different server. Given $j_t$, the $t$-th server computes $C(x)$ and returns that $j_t$-th entry of $C(x)$. Given these answers, the user completes the simulation of $A^{C(x)}(i)$ and computes $x_i$.

This PIR system is 1-secure, has perfect recovery, the query size is $\log n$ and the answer size is $\log |\Gamma|$.

The main known results for constant values of $q$ are shown in Table 1.

For $q = 2$, the Hadamard code gives a perfectly smooth code with exponential encoding length. The exponential blow-up is necessary for all 2-query binary LDCs [KdW03]. Even for $q = 2$ and large alphabets, tight results are not known. Chor et al. [CGKS98] show that one can achieve encoding length roughly $2^{k^{1/3}}$ with an alphabet of size roughly $2^{k^{1/3}}$, which corresponds to a PIR with communication complexity $O(k^{1/3})$. For such alphabet sizes, the only applicable lower bounds are in [KT00], and they are barely super-linear.

For $q = 3$, even for binary alphabet there is a huge gap between a polynomial lower bound and a sligthly sub-exponential upper bound. For larger $q$, the best known Boolean constructions achieve encoding length just slightly better than $2^{k^{1/q}}$, while the lower bound is roughly $k^{1+1/q}$.

Below we give some more references to constructions and lower bounds.

## 3.3 Local Decoders with Constant Query Complexity

We give an overview of some simple constructions of perfectly smooth codes. We will not get into details of the best known construction, the one by Beimel et al [BIKR02], which is somewhat complicated.

---

[15] Recall that all the oracle algorithms in this paper make non-adaptive queries.

### 3.3.1 Hadamard Code

Let us start by considering the Hadamard code $H : \{0,1\}^k \to \{0,1\}^{2^k}$. In the Hadamard code, the encoding $H(x)$ has one entry for every vector $a \in \{0,1\}^k$, and the content of the $a$-th location of $H(x)_a$ is the bit $a \cdot x = \sum_j a_j x_j \pmod 2$. Suppose we have oracle access to $H(x)$ and we want to reconstruct $x_i$ for some index $i$. Clearly $x_i = e_i \cdot x$, where $e_i$ is a vector with a 1 in the $i$-th position and 0s elsewhere, however we cannot just read the $e_i$-th position of $H(x)$, because a smooth decoder must make uniformly distributed queries. Instead, we use the following trick: we pick at random a vector $a \in \{0,1\}^k$ and we read the entries $H(x)_a$ and $H(x)_{a \oplus e_i}$, which will return, respectively, $a \cdot x$ and $(a \oplus e_i) \cdot x$. By linearity, the xor of these two values will be $((a \oplus a \oplus e_i) \cdot x) = e_i \cdot x = e_i$. This idea dates back to [BLR93].

If the smooth decoder is applied to a string $y$ that is at relative distance $\delta < 1/4$ from a valid codeword $C(x)$, then, for every $i$, the decoder succeeds with probability at least $1 - 2\delta$ in computing $x_i$. If the relative distance between $C(x)$ and $y$ is $1/4 - \epsilon$, then the success probability of the decoder is $1/2 + 2\epsilon$, which can be amplified to, say, $1 - 1/4k$ by repeating the decoding independently for $O(\epsilon^{-2} \cdot \log k)$ times and then taking the majority value. If we do this for every $i$, we get an algorithm that runs in time $O(\epsilon^{-2} k \log k)$ and computes $x$ with probability at least $3/4$.

### 3.3.2 Polynomial Codes

A similar idea (dating back to [BF90]) can be used to get a perfectly smooth decoder for a variant of the Reed-Muller code. The variant will have the property of being a *systematic* code, meaning that the message occurs as a substring of the encoding.

To make the Reed-Muller code systematic, we proceed as follows. We have a field $\mathbb{F}$, a subset $S \subseteq \mathbb{F}$, a degree parameter $t$ and a number of variables $m$. Besides, we also choose another subset $A \subseteq S$ of size $|A| = t/m$. In the standard Reed-Muller code, we would just consider all possible $\binom{m+t}{m}$ coefficients that an $m$-variate degree-$t$ polynomial can have, and we encode a message of length $\binom{m+t}{m}$ by interpreting each message coordinate as a coefficient. This time, instead, we have a (shorter) message of length $|A|^m = (t/m)^m$, which we think of as a function $f : A^m \to \mathbb{F}$; using interpolatio, we find a polynomial $p$ of degree $\leq t$ that agrees with $f$ on $A^m$. The evaluation of this polynomial on $S^m$ will be our encoding.

This is not much worse than the standard encoding, and, for example, it remains true that the rate is constant for constant $m$. The advantage, is that the original message is a subset of the encoding. (The evaluation of the polynomial at $A^m$.) A code with such a property is called a *systematic* code.

In the following discussion we assume $S = \mathbb{F}$.

Consider now the following algorithm that, given oracle access to a polynomial $p : \mathbb{F}^m \to \mathbb{F}$ of degree $t$, and given $a$, computes $p(a)$ by making random oracle queries.

1. Choose a random $b \in \mathbb{F}^m$ and consider the line $l(z) = a + zb$.

2. Query $p$ at locations $l(1), l(2), \ldots, l(t+1)$.

3. Compute the unique univariate degree $t$ polynomial $q$ such that $q(z) = p(l(z))$ for $z = 1, 2, \ldots, t+1$. Return $q(0)$.

Where, in the algorithm description, we used "2," ... "$t+1$" as the names of $t$ elements of $\mathbb{F}$ distinct from 1 and 0.

The algorithm is based on the following idea: if $p(x_1,\ldots,x_m)$ is a multivariate polynomial of degree $t$, and $l(z) = (a_1 + zb_1,\ldots,a_m + zb_m)$ is a line, then $P(l(z))$ is a univariate polynomial in $z$ of degree at most $t$. As such, it can be reconstructed if we are given its values at $t+1$ different points.

So we pick at random a line $l()$ that passes through the point $a$ (at which we want to evaluate $p()$), we read the value of $p(l()$ at $t+1$ different points, we reconstruct the univariate polynomial $p(l(z))$, and finally, by evaluating it at 0, we compute $p(l(0)) = p(a)$. The second observation to make, which completes the analysis, is that, on a random line of the form $l(z) = a + zb$ every point, except $l(0) = a$, is uniformly distributed, and so the procedure is making uniform queries.

A disadvantage of this procedure is that it can never lead to a constant query complexity, because the degree can never be a constant in the version of the Reed-Muller code we described before. We now describe another variant in which we encode a message as a constant-degree polynomial, so that the above algorithm has constant query complexity.

Suppose that $k \leq \binom{m}{t}$. We index each message entry by a unique $d$ element subset $S$ of $[m]$. Given a message $x = (x_S)_{S \subseteq [m]}$, we define the polynomial

$$p_x(z_1,\ldots,z_m) = \sum_{S:|S|=d} x_S \prod_{j \in S} z_j$$

over the field $\mathbb{F}$, where $|\mathbb{F}| < 2t$. The encoding $C(x)$ is obtained by evaluating $p_x$ over all points in $\mathbb{F}^m$.

Since $p_x$ has total degree $t$, the decoding algorithm described above also works for the code $C$. Moreover, $C$ is systematic: We can write $x_S = p_x(e_S)$, where $e_S = (e_{S1},\ldots,e_{Sm}) \in \mathbb{F}^m$ and

$$e_{Sj} = \begin{cases} 1 & \text{if } j \in S, \\ 0 & \text{otherwise.} \end{cases}$$

The code $C$ can encode messages of length up to $\binom{m}{t}$ and has codeword length $n = |\mathbb{F}|^m$, provided $|\mathbb{F}| > t$. If we take, say, $|\mathbb{F}| \leq 2t$, this yields $n = 2^{O(t \log t k^{1/t})}$, where $t = q - 1$.

### 3.4 Local Decoder with Polylogarithmic Complexity

If we want decoders with constant query complexity, then the best known constructions are super-polynomial.

By adjusting the parameters in the Reed-Muller-like constructions of the previous sections, it is however possible to have perfectly smooth codes with $n = \text{poly}(k)$, or even $n = O(k)$, and sub-linear complexity. The problem is that a perfectly smooth decoder of super-constant query complexity is not necessarily a good local decoder able to correct a large number of errors.

There is, however, a way to get a good local decoder even when the degree of the polynomial (and so the query complexity of the decoding algorithm) are high.

We are considering the following problem: we have a message which we view as a function $f : A^m \to \mathbb{F}$, for some subset $A \subseteq \mathbb{F}$. We encode the message by finding a polynomial $p$ of degree $t$ that agrees with $f$ on all of $A^m$, and the encoding is the evaluation of $p$ at all points of $\mathbb{F}^m$.

Now we have oracle access to a corrupted encoding $g : \mathbb{F}^m \to \mathbb{F}$, which disagress from $p$ in a $\delta$ fraction of elements of $\mathbb{F}^m$. We are given an entry $a \in A^m$ and we would like to compute $f(a) = p(a)$ with high probability by using oracle access to $g$.

As before, we pick a random $b \in \mathbb{F}^m$ and consider the random line $l(z) = a + bz$, $z \in \mathbb{F}$, and we would like to find the univariate polynomial $q(z) = p(l(z))$, because if we find $q()$ then we have also found $f(a) = p(a) = q(0)$. Instead of just reading $t+1$ points of $p(l(z))$ and interpolating (we don't have access to $p()$), we read $g(l(z))$ for, say, $3t$ values of $z$, and then we apply the Berlekamp-Welch algorithm to find the degree-$t$ polynomial $q()$ that agrees the most with these values. The Berlekamp-Welch algorithm will indeed find $q()$ provided that there are fewer than $t$ places among those we read where $g(l())$ differs from $p(l())$. Recall that each of these places is a point on a line, and so it is random, so that on average we expect to encounter at most $3\delta t$ errors, which is less than $t/4$, say, if $\delta < 1/12$. If so, then a Markov argument shows that with probability at least $3/4$ we indeed find fewer than $t$ errors and so we correctly find $q()$.

This is far from optimal, but it gives an idea of how good LDCs with super-constant query complexity work.

By playing with the parameters, it is possible to construct, for every $\epsilon$, binary LDCs $C : \{0,1\}^k \to \{0,1\}^n$ where $n = O(k)$ and the decoding algorithm has query complexity $n^\epsilon$. The best lower bound on query complexity for LDCs of linear encoding length is logarithmic, and so we have again an exponential gap between constructions and lower bounds. In this case too my guess is that the lower bound should be improved.

**Open Question 1** *Prove that there cannot be a binary $(q, \delta, 3/4)$-LDC $C : \{0,1\}^k \to \{0,1\}^n$ where $q = O(\log n)$ and $n = O(k)$.*

## 3.5 Application to Average-Case Complexity

Techniques sketched in the previous section show the existence of LDCs $C : \{0,1\}^k \to \{0,1\}^n$ with $n = \text{poly}\,k$, with a decoder of poly log $k$ complexity, and with a polynomial time encoding algorithm.

Let us now see how this applies to average-case complexity.

Let $L$ be an EXP-complete problem, and for an input length $t$ let us consider the restriction of $L$ to inputs of length $L$. We can see $L$ restricted to some inputs as being a binary string of length $2^t$. (The truth-table of a Boolean function with $t$-bits input.)

Let us encode this string using our code $C$: we get a string of length $2^{O(t)} = 2^{t'}$, and let us think of this string as defining a new problem $L'$ on inputs of length $t'$. If $L$ was in EXP, then so is $L'$. In fact, if $L$ was EXP-complete, then so is $L'$. In addition, if we have an algorithm for $L'$ that is good on average, the algorithm together with the local decoder give a probabilistic algorithm for $L$ that works on all input, and EXP $\subseteq$ BPP.

This argument shows that if every problem in EXP can be solved well on average then EXP $\subseteq$ BPP.

This argument raises many questions, such as the following:

- Can the same argument work for PSPACE? The answer is yes, provided we can construct a good LDC in logarithmic space. This is in fact easy to do, and so the same argument does indeed apply to PSPACE.

- Can the same argument work for NP? Viola [Vio03] shows that the LDC decoding does not work for NP, and, more generally, for any problem in polynomial hierarchy. Bogdanov and Trevisan [BT03] show that worst-case to average-case reductions in NP are problematic even if they are not based on LDCs.

## 3.6 Lower Bounds

In this section we give an overview of lower bounds for locally decodable codes.

The following notion is a useful relaxation of the notion of perfectly smooth code.

**Definition 4 (Smooth Decoder)** *A $(q, c, p)$-smooth decoder for a code $C : \Sigma^k \to \Gamma^n$ is a probabilistic oracle algorithm A of query complexity q such that for every $i \in k$ and every $x \in \{0, 1\}^k$ we have*
$$\mathbf{Pr}[A^{C(x)}(i) = x_i] \geq p$$
*Furthermore, for every $t \in [q]$, every $i \in [k]$, and every $j \in [n]$, the probability that the t-th query of $A^{(\cdot)}(i)$ is j is at most $c/n$. A code having a $(q, c, p)$-smooth decoder is also called a $(q, c, p)$-smooth code.*

A perfectly smooth decoder of query complexity $q$ is a $(q, 1, 1)$-smooth decoder.

The definition of smooth decoder allows the algorithm to use non-uniform distributions of queries, however no query has a probability of being made which is much higher than with respect to the uniform distribution.

The following result shows that every locally decodable code has a smooth decoder.

**Lemma 9 ([KT00])** *If $C : \Sigma^k \to \Gamma^n$ is a $(q, \delta, p)$-LDC, then it is also a $(q, 1/\delta, p)$-smooth code.*

Goldreich et al. [GKST02] also show that every private information retrieval system implies a smooth code with related parameters.

These results show that it is enough to prove lower bounds for smooth codes, a strategy followed explicitely in [KT00, GKST02] and implicitly in [KdW03].

Another step is the following.

**Lemma 10** *If $C : \Sigma^k \to \Gamma^n$ is a $(q, c, 1/|\Sigma| + \epsilon)$-smooth code, then for every index $i \in [k]$ there is a collection $M_i$ of $\Omega(n\epsilon/c)$ disjoint q-tuples $(j_1, \ldots, j_q)$ such that for a random x it is possible to predict with probability bounded away form half the entry $x_i$ given the entries $C(x)_{j_1}, \ldots, C(x)_{j_q}$.*

If $C : \mathbb{F}^k \to \mathbb{F}^n$ is linear, then the conclusion of the above Lemma has a simpler interpretation. If we write $C(x) = (c_1 \cdot x, \ldots, c_n \cdot x)$, then for every $i$ there is a a collection $M_i$ of $\Omega(n\epsilon/c)$ disjoint $q$-tuples $(j_1, \ldots, j_q)$ such that the vector $e_i$ is in the span of $c_{j_1}, \ldots, c_{j_q}$.

A converse can also be proved, that is, if the above collections exist then the code is smooth and also locally decodable, which means that there is no loss in generality in trying to prove a lower bound using this combinatorial structure. The exponential lower bound in [GKST02] follows then from the following combinatorial result.

**Lemma 11** *Let $a_1, \ldots, a_n \in \{0, 1\}^k$ be (not necessarily distinct) vectors and suppose that for every $i \in [k]$ there is a set $M_i$ of $\delta n$ disjoint pairs $(j, j')$ such that $a_j \oplus a_{j'} = e_i$.*

*Then $n \geq 2^{2\delta k}$.*

A proof of the following combinatorial result would give the first super-polynomial lower bound for the case of linear codes decodable with three queries. (So far, super-polynomial lower bounds have been proved only for the case of two queries. Upper bounds are super-polynomials for all constant query complexities.)

**Open Question 2** *Suppose that $a_1, \ldots, a_n$ is a sequence of elements of $\{0,1\}^k$ such that for every $i \in [k]$ there is a set of $\Omega(n)$ disjoint triples $(j_1, j_2, j_3)$ such that $a_{j_1} \oplus a_{j_2} \oplus a_{j_3} = e_i$. Prove that $n$ must be super-polynomial in $k$.*

This is the natural next lower bound question to adress, and it is much harder than it looks.

The best lower bounds for LDCs are currently those from [KdW03], proved using quantum information theory.

The use of quantum arguments in a "classical" problem is surprising, and it would be interesting to see a purely combinatorial proof of the same results.

**Open Question 3** *Re-prove the results of [KdW03] without using quantum information theory.*

## 3.7 Notes and References

The notion of locally decodable codes was implicitly discussed in various places in the early 1990s, most notably in [BFLS91, Sud92], and it was explicitely defined in [KT00], where smooth codes are also defined. The notion of private information retrieval was introduced by Chor and others [CGKS98]. Locally decodable codes, private information retrieval and smooth codes can be seen as the combinatorial analogs of notions that had been studied in complexity theory in the late 1980s and early 1990s. In particular, one can see the decoding procedure of a locally decodable codes as a combinatorial version of a self-corrector [BK89, Lip90, BLR93], a perfectly smooth decoder is analogous to a random-self-reduction, a notion explicitely defined in [AFK89, FKN90], and a private information retrieval system is analogous to an "instance-hiding" scheme [AFK89].

The perfectly smooth decoder of Hadamard Codes is due to Blum and others [BLR93] and the Reed-Muller codes is due to Beaver and Feigenbaum [BF90]. There has been a substantial amount of work devoted to the construction of efficient private information retrieval schemes, leading to the sofisticated construction of Beimel and others [BIKR02]. Work by Ambainis, Beimel, Ishai and Kushilevitz [Amb97, IK99, BI01] is particularly notable. The Reed-Muller decoder of Section 3.4 is due to Gemmell and others [GLR$^+$91]. It should be noted that there are other models and questions about private informational retrieval that we did not discuss in this section. In particular, we did not discuss the notion of *computationally* secure private information retrieval, in which the distribution of queries are just computationally indistinguishable, the notion of *symmetric* private information retrieval, in which the user does not get any other information about $x$ except $x_i$, and other algorithmic problems, such as issues of efficiency for the server.

Lower bounds for private information retrieval were first proved by Mann [Man98]. Lower bounds for smooth codes, which imply lower bounds for locally decodable codes and for private informational retrieval, are proved in [KT00, GKST02, Oba02, KdW03].

## 4 Sublinear Time List Decoding

Sub-linear time list-decoding is perhaps a less intuitive notion to define than sub-linear time unique decoding.

As in the past section, we have some code $C : \{0,1\}^k \to \{0,1\}^n$ and we have oracle access to a string $y \in \{0,1\}^n$ that is within some distance $d$ from a codeword, and we want to find all the messages $x$ such that $C(x)$ and $y$ are within distance $d$.

Since the decoder has to run in time $o(k)$, it cannot output the full list, but rather it will output a list of "compressed" representations of messages. What will a compressed representation of a message $x$ be like: it will be the code of an efficient probabilistic oracle algorithm that given $i$ and oracle access to $y$ gives $x_i$ in output with high probability.

Another way to look at this setting is to think of the decoding algorithm as outputting a list of "local decoders" as previously defined, one for every message in the list.

This model is discussed in detail in [Sud00], along with the description of several applications. We will quickly review such applications, and refer the reader to [Sud00] for more details, and we will devote more space to results by Akavia et al [AGS03], that postdate [Sud00].

## 4.1 Formal Definition of Local List-Decoder

Let us fix a model of computation to describe oracle algorithms.

**Definition 5 (Local List-Decoding)** *A probabilistic oracle algorithm $A$ is a local list-decoder for a code $C : \Sigma^k \to \Sigma^n$ for radius $r$ if, for every string $y \in \Sigma^n$, $A^y$ outputs a list of probabilistic oracle algorithms $D_1, \ldots, D_L$ such that for every string $x$ such that $d_H(C(x), y) \leq rn$ the following happens with probability at least $3/4$ over the random choices of $A^y$:*

$$\exists j \in [L]. \forall i \in [k]. \mathbf{Pr}[D_j^y(i) = x_i] \geq 3/4 .$$

*The probability in the above expression is taken over the random choices of $D_j^y$.*

We note that we are interested in the complexity both of $A$ and of $D_1, \ldots, D_L$, and, ideally, all these complexities would be a constant. Interestingly, constant complexity (at least, constant oracle query complexity) is achievable for the Hadamard code when $r = 1/2 + \epsilon$, for constant $\epsilon$.

In applications to average-case complexity and cryptography, running time poly-logarithmic in $n$ is also acceptable, and we will sketch a proof (due to Sudan and others [STV01] of the fact that a veriant of the Reed-Muller codes have local list decoder of polylogarithmic complexity in $k$ and $n$ and polynomial in $1/r$.

In the case of the Hadamard code, for which $n = 2^k$, poly-logarithmic complexity in $n$ is equivalent to polynomial complexity in $k$, and a local list-decoder has enough time to explicitly output the list of codewords. Goldreich and Levin [GL89] present a list-decoder for the Hadamard code that runs in time polynomial in $k$ and in $1/\epsilon$ when $r = 1/2 + \epsilon$. The local decoder of constant (depending on $\epsilon$) query complexity can be derived from an alternative proof of the result of Goldreich and Levin attributed to Rackoff. We note that in, most applications of the Goldreich-Levin result, $1/\epsilon$ is either polynomially related to $k$, or it is even superpolynomial in $k$, a local decoder of complexity polynomial in $1/\epsilon$ and independent of $k$ is not significantly more efficient than the original Goldreich-Levin algorithm.

## 4.2 Local List Decoders for the Hadamard Code and for Polynomial Codes

We state the some known results about local list-decoding of error-correcting codes. We will prove the Goldreich-Levin result in Section 4.4.

**Theorem 12 ([GL89])** *Let $H : \{0,1\}^k \to \{0,1\}^{2^k}$ be the Hadamard code. There is an algorithm that, given oracle access to $y \in \{0,1\}^{2^k}$ and a parameter $\epsilon$, runs in $\text{poly}(k/\epsilon)$ time and outputs,* with high probability, a list of all the strings $x \in \{0,1\}^k$ such the Hamming distance between $y$ and $C(x)$ is at most $1/2 - \epsilon$.

The theorem has a stronger form in which the algorithm runs in poly$(1/\epsilon)$ time, independent of $\epsilon$, and outputs a list of local decoders, each running in poly$(1/\epsilon)$ time.

The next result that we state is for polynomial encodings.

Let $\mathbb{F}$ be a field, $m$ be an integer, $A \subseteq \mathbb{F}$ be a subset. Consider the polynomial encoding $C : \mathbb{F}^{|A|^m} \to \mathbb{F}^{|\mathbb{F}|^m}$ in which we think of a message as a function $f : A^m \to \mathbb{F}$, and its encoding is obtained by finding, using interpolation, a polynomial $p$ of degree $\leq m|A|$ that agrees with $f$ on $A^m$, and then evaluating $p$ on all the points of $\mathbb{F}^m$.

**Theorem 13 ([STV01])** *There is a constant $c$ such that the following happens. Let $C : F^{|A|^m} \to \mathbb{F}^{|F|^m}$ be the polynomial encoding described above. There is an algorithm that, given oracle access to a string $g \in \mathbb{F}^{|F|^m}$ and a parameter $\epsilon > c \cdot \sqrt{|A|m/|\mathbb{F}|}$, runs in time poly$(\epsilon^{-1}, m, |S|, \log|\mathbb{F}|)$ and outputs a list of local decoders such that, for every $f \in \mathbb{F}^{|S|^m}$ such that the relative distance between $P(f)$ and $g$ is less than $1 - \epsilon$, there is a local decoder in the list that computes $f$ given oracle access to $g$.*

In simpler terms, Theorem 13 states that there for every $\epsilon > 0$ there is an efficiently computable error correcting code $C : \mathbb{F}^k \to \mathbb{F}^n$ such that $n = \text{poly}(\epsilon^{-1}, k)$ and such that local list decoding can be performed in poly$(\epsilon^{-1}, \log(n))$ time even after $(1 - \epsilon)n$ errors occurred.

The code can be concatenated with a binary code to obtain a binary locally list-decodable code.

**Theorem 14 ([STV01])** *For very $\epsilon$ and $k$ there is a polynomial time encodable code $C : \{0,1\}^k \to \{0,1\}^n$ with $n = \text{poly}(k, \epsilon^{-1})$ and a local list decoding algorithm that runs in time poly$(\epsilon^{-1}, \log n)$ and is able to correct up to $(1/2 - \epsilon)n$ errors.*

## 4.3 Applications to Average Case Complexity

Average-case complexity is an ideal application for sub-linear time list decoder, since one can deal with average-case algorithms that make a very large fraction of errors. On the other hand, coding-theoretic methods can prove average-case complexity results only for classes like PSPACE and EXP, while one is typically interested in the average-case complexity of problems within NP.

A strong motivation to the study of average-case complexity in EXP came from a result by Nisan and Wigderson [NW94]. Before stating the result, let us introduce the following notion: a decision problem on inputs of length $n$ is $(S(n), \delta(n))$-average case hard if every circuit $C$ of size $\leq S(n)$ fails to solve the problem on at least a $\delta(n)$ fraction of inputs of length $n$.

**Theorem 15 (Nisan-Wigderson)** *Suppose that there is a problem in* DTIME$(2^{O(n)})$ *that is* $(2^{\Omega(n)}, 1/2^{\Omega(n)})$*-average case hard. Then* P = BPP.

The Nisan-Wigderson result shows an extremely strong conclusion from a very strong assumption. The postulated average-case complexity is very high, and the assumption would sound more natural if it referred to standard (worst-case) circuit complexity.

**Theorem 16 (Impagliazzo Wigderson [IW97])** *Suppose that there is a problem $L$ in* DTIME$(2^{O(n)})$ *that has circuit complexity $2^{\Omega(n)}$; then there is also a problem $L'$ in* DTIME$(2^{O(n)})$ *that is $(2^{\Omega(n)}, 1/2^{\Omega(n)})$-average case hard. (And* P = BPP*.)*

The Impagliazzo-Wigderson proof was very complicated, and their construction included several parts, one of them being, essentially, a Reed-Muller encoding of the starting problem.

The code and the decoder of Theorem 14 give a simpler proof of Theorem 16, as follows. Let $L$ be problem in the assumption of Theorem 16. For everyy $n$, let us consider the binary string $x$ of length $K = 2^n$ that describes the "truth-table" of $L$ for inputs of length $n$. Let us compute $C(x)$, which is of length $N = \text{poly}(K) = 2^{cn}$ for some constant $c$. We define $L'$ to be the problem whose truth-table, on inputs of length $cn$, is $C(x)$. Having a circuit that computes $L'$ correctly on a $1/2 + \epsilon$ fraction of inputs is essentially the same as having oracle access to a string of length $N$ that is within distance $1/2 - \epsilon$ from $C(x)$. Applying the linear decoder, we can find a list of programs such that one of them computes $x$ on any entry of our choice (that is, it solves $L$ on any input of length $n$ of our choice) probabilistically in time polynomial in $1/\epsilon$ and in $n$. Since our goal is to build a circuit, we can non-uniformly pick the correct program from the list, and convert the probabilistic algorithm into a deterministic circuit. If we had a circuit of size $2^{\delta cn}$ that computed $L'$ on a $1/2 + 1/2^{\delta cn}$ fraction of inputs of length $cn$, and if we picked $\epsilon$ appropriately, we end up constructing a circuit of size $2^{O(\delta n)}$ that solves $L$ on all inputs of length $n$, which contradicts the assumption of the Theorem if $\delta$ is small enough.

## 4.4 Proof of the Goldreich-Levin Result

In this discuss the Goldreich-Levin list-decoding algorithm for the Hadamard code. It will be convenient to think of the codewords of the Hadamard code as functions:

**Definition 6** *Let $a \in \{0,1\}^k$, and define $L_a : \{0,1\}^k \to \{0,1\}$ to be the function $L_a(x) = a \cdot x$. Then, $L_a(\cdot)$ is the Hadamard encoding of $a$.*

We may then state the main result from [GL89] as follows.

**Theorem 17 ([GL89])** *There is a (probabilistic) algorithm that given oracle access to a function $g : \{0,1\}^k \to \{0,1\}$ and a parameter $\epsilon > 0$ runs in time $O\left(\frac{1}{\epsilon^4} k \log k\right)$ and outputs a list of $O\left(\frac{1}{\epsilon^2}\right)$ elements of $\{0,1\}^k$ such that: for every $a$ for which $L_a$ and $g$ agree on $> \frac{1}{2} + \epsilon$ fraction of inputs, the probability that $a$ is in the list is at least $3/4$.*

We recall from Section 3.3.1 that if we are given an oracle that agrees with a linear function $L_a$ on, say, a $7/8$ fraction of the inputs, then it is possible to compute $a$ in time $O(k \log k)$. Our goal will be to able to simulate an oracle that has good agreement with $L_a$ by "guessing" the value of $L_a$ at a few points.

We first choose $t$ random points $x_1 \ldots x_t \in \{0,1\}^n$ where $t = O(1/\epsilon^2)$. For the moment, let us suppose that we have "magically" obtained the values $L_a(x_1), \ldots, L_a(x_k)$. Then define $g'(z)$ as the majority value of:

$$L_a(x_j) \oplus g(z \oplus x_j) \qquad j = 1, 2, \ldots, t \tag{1}$$

Since for each $j$ we obtain $L_a(z)$ with probability at least $\frac{1}{2} + \epsilon$, by choosing $t = O(1/\epsilon^2)$ we can ensure that

$$\mathbf{Pr}_{z, x_1, \ldots, x_t}\left[g'(z) = L_a(z)\right] \geq \frac{31}{32}. \tag{2}$$

from which it follows that

$$\mathbf{Pr}_{x_1, \ldots, x_k}\left[\mathbf{Pr}_z\left[g'(z) = L_a(z)\right] \geq 7/8\right] \geq \frac{3}{4}. \tag{3}$$

Consider the following algorithm.

**Algorithm** GL-First-Attempt:
  **pick** $x_1, \ldots, x_t \in \{0,1\}^k$ where $t = O(1/\epsilon^2)$
  **for all** $b_1, \ldots, b_t \in \{0,1\}$
    **define** $g'_{b_1 \ldots b_t}(z)$ as majority of: $b_j \oplus g(z + x_j)$
    **apply** the algorithm of Section 3.3.1 to uniquely decode $g'_{b_1 \ldots b_t}$
    **add** result to list

The idea behind this program is that we do not in fact know the values $L_a(x_j)$, so we guess all possibilities by considering all choices for the bits $b_j$. For each $a$ such that $L_a$ and $g$ agree on more than half of their domain, we will eventually choose $b_i = L_a(x_j)$ for all $j$ and then, with high probability, recover $a$ via the algorithm of Section 3.3.1. The obvious problem with this algorithm is that its running time is exponential in $t = O(1/\epsilon^2)$ and the resulting list may also be exponentially larger than the $O(1/\epsilon^2)$ bound promised by Theorem 17.

To overcome these problems, consider the following similar algorithm.

**Algorithm GL:**
  **choose** $x_1, \ldots, x_l \in \{0,1\}^k$ where $l = O(\log(1/\epsilon))$
  **for all** $b_1, \ldots, b_l \in \{0,1\}$
    **define** $g'_{b_1 \ldots b_l}(z)$ as majority over all nonempty $S \subseteq \{1, \ldots, l\}$ of: $(\oplus_{j \in S} b_j) \oplus g\left(z + \sum_{j \in S} x_j\right)$
    **apply** the algorithm of Section 3.3.1 to uniquely decode $g'_{b_1 \ldots b_l}$
    **add** result to list

Let us now see why this algorithm works. First we define, for any nonempty $S \subseteq \{1, \ldots, l\}$, $x_S \equiv \sum_{j \in S} x_j$. Then, since $x_1, \ldots, x_l \in \{0,1\}^k$ are random, it follows that for any $S \neq T$, $x_S$ and $x_T$ are independent and uniformly distributed. Now consider any $a$ such that $L_a(x)$ and $g(x)$ agree on $\frac{1}{2} + \epsilon$ of the values in their domain. Then for the choice of $\{b_j\}$ where $b_j = L_a(x_j)$ for all $j$, we have that

$$\bigoplus_{j \in S} b_j = L_a(x_S)$$

and, with probability $\frac{1}{2} + \epsilon$,

$$g\left(z \oplus \sum_{j \in S} x_j\right) = g(z \oplus x_S) = L_a(z \oplus x_s) = L_a(z) \oplus L_a(x_S)$$

so combining the above results yields

$$\bigoplus_{j \in S} b_j \oplus g\left(z \oplus \sum_{j \in S} x_j\right) = L_a(z)$$

with probability $\frac{1}{2} + \epsilon$.

Note the following simple lemma whose proof we omit:

**Lemma 18** *Let $R_1, \ldots, R_t$ be a set of pairwise independent $0-1$ random variables, each of which is 1 with probability at least $\frac{1}{2} + \epsilon$. Then $\mathbf{Pr}[\sum_i R_i \geq t/2] \geq 1 - O(\frac{1}{\epsilon^2 t})$.*

Lemma 18 allows us to upper-bound the probability that the majority operation used to compute $g'$ gives the wrong answer. Combining this with our earlier observation that the $\{x_S\}$ are pairwise independent, we see that choosing $l = 2\log 1/\epsilon + O(1)$ suffices to ensure that $g'_{b_1 \ldots b_l}(z) = L_a(z)$ with probability $1 - c \geq \frac{3}{4} + \epsilon$ for any constant $c > 0$. Thus we can use the algorithm of Section 3.3.1 to obtain $a$ with high probability. Choosing $l$ as above ensures that the list generated is of length at most $2^l = O(1/\epsilon^2)$ and the running time is then $O(k \log k/\epsilon^4)$, due to the $O(1/\epsilon^2)$ iterations of the algorithm of Section 3.3.1. This completes the proof of the Goldreich-Levin theorem.

## 4.5 Applications to "Hard-Core Predicates"

### 4.5.1 Hard-core Predicates for One-way Permutations

Intuitively, a function $f : \{0,1\}^k \to \{0,1\}^k$ is a one-way permutation if it is easy to compute but hard on average to invert. Note that $f$ is a permutation on the set $\{0,1\}^k$, and not a function that permutes the bits of its input string. Formally,

**Definition 7** *A permutation $f : \{0,1\}^k \to \{0,1\}^k$ is $(s, \epsilon)$-one-way if*

- *there is an efficient algorithm that on input $x$ outputs $f(x)$*
- *for all circuits $D : \{0,1\}^k \to \{0,1\}^k$ of size $s$, and for all sufficiently large $k$,*

$$\Pr_x[D(f(x)) = x] \leq \epsilon$$

Often in applications such as the construction of pseudorandom generators in cryptography, we are interested in a hard-on-average decision problem. This motivates the notion of a hard-core predicate:

**Definition 8** *A function $B : \{0,1\}^k \to \{0,1\}$ is a $(s, \epsilon)$ hard-core predicate for a permutation $f$ if it is hard on average to compute $B(x)$ given $f(x)$, that is: for all circuits $D : \{0,1\}^k \to \{0,1\}$ of size $s$, and for all sufficiently large $k$,*

$$\Pr_x[D(f(x)) = B(x)] \leq 1/2 + \epsilon$$

### 4.5.2 Hard-core Predicates Using Goldreich-Levin

The Goldreich-Levin theorem gives us a hard-core predicate for any one-way permutation (with a slight modification) whose hardness is closely related to that of the original permutation:

**Theorem 19** *If $f$ is a $(s, 1/s)$-one-way, then $B(x, r) = x \cdot r$ is a $(s^{\Omega(1)}, 1/s^{\Omega(1)})$ hard-core predicate for the permutation $f' : \{0,1\}^{2k} \to \{0,1\}^{2k}$, where $f'(x, r) = (f(x), r)$.*

PROOF: The proof is by contradiction: given an algorithm $A$ that computes the hard-core predicate $B$ on a $1/2 + \epsilon$ fraction of its inputs, we produce an algorithm that inverts $f$ on some $O(\epsilon)$ fraction of inputs, contradicting the one-way-ness of $f$ for a suitable choice of parameters. The reduction is uniform, so we will present the proof for a uniform algorithm $A$; the result extends readily to circuits. More precisely, having a uniform reduction means that if we start with a one-way permutation that is hard to invert using polynomial-time algorithms (respectively polynomial-sized circuits), we obtain a hard-core predicate that is hard to predict using polynomial-time algorithms (respectively polynomial-sized circuits).

Now, suppose we are given an algorithm $A$ that runs in time $t$ (or a circuit of size $t$ for the non-uniform setting) such that:

$$\Pr_{x,r}[A(f(x), r) = x \cdot r] \geq 1/2 + \epsilon$$

Then by an averaging argument, we have:

$$\Pr_x[\Pr_r[A(f(x), r) = x \cdot r] \geq 1/2 + \epsilon/2] \geq \epsilon/2$$

Fix any $x$ such that

$$\Pr_r[A(f(x), r) = x \cdot r] \geq 1/2 + \epsilon/2$$

Note that as we vary $r$ over $\{0,1\}^k$, $x \cdot r$ yields the Hadamard encoding of $x$ and therefore $A(f(x), \cdot)$ yields a good approximation to this encoding. We can then recover a list of candidates for $x$ via list-decoding. The precise reduction is as follows: On input $f(x)$,

1. Define $g(r) = A(f(x), r)$. For an $\epsilon/2$ fraction of the choices of $x$, $g(r)$ and $L_x(r)$ agree on $1/2 + \epsilon/2$ fraction over the choices of $r$.

2. Run the Goldreich-Levin algorithm using $g(\cdot)$ as an oracle, and with parameter $\epsilon/2$. This takes time $O\left(\frac{1}{\epsilon^4} k \log k\right)$ and computes a list of size $O\left(\frac{1}{\epsilon^2}\right)$.

3. For each element $x'$ of the list, output $x'$ if $f(x') = f(x)$.

For a $\epsilon/2$ fraction of the choices of $x$, this algorithm outputs $x$ on input $f(x)$ with probability $3/4$. This yields an algorithm $A'$ that runs in time $O\left(t \cdot \frac{1}{\epsilon^4} k \log k + \frac{1}{\epsilon^2} k^{O(1)}\right)$ satisfying

$$\Pr_x[A(f(x)) = x] \geq 3\epsilon/8$$

which means $f$ cannot be $(poly(k, 1/\epsilon, t), O(\epsilon))$-one-way. It follows that if $f$ is $(s, 1/s)$-one-way, then $B$ is a $(s^{\Omega(1)}, 1/s^{\Omega(1)})$ hard-core predicate for the permutation $f'$. $\square$

**Remark 1** *Note that if $f$ is one-way, then $f'$ is also one-way.*

### 4.5.3 Hard-core Predicates via (Efficiently) List-Decodable Codes

Let $C : \{0,1\}^k \to \{0,1\}^n$ be a binary code, and $f : \{0,1\}^k \to \{0,1\}^k$ be a permutation. Consider the predicate $B : \{0,1\}^k \times [n] \to \{0,1\}$ given by $B(x,j) = C(x)_j$ and the corresponding function $f' : \{0,1\}^k \times [n] \to \{0,1\}^k \times [n]$ given by $f'(x,j) = (f(x), j)$. Extending the result in the previous section (which corresponds to $C$ being the Hadamard code), it is easy to see that if $f$ is a one-way permutation, and $C$ is an efficiently list-decodable code, then $B(x,j) = C(x)_j$ is a hard-core predicate for $f'$. The reduction is as follows:

Suppose we are given an algorithm $A$ that runs in time $t$ such that:

$$\Pr_{x,j}[A(f(x),j) = C(x)_j] \geq 1/2 + \epsilon$$

Then,

$$\Pr_{x}[\Pr_{j}[A(f(x),j) = C(x)_j] \geq 1/2 + \epsilon/2] \geq \epsilon/2$$

Now, given $f(x)$, use the list-decoding for $C$ to find a list of all codewords with agreement at least $1/2 + \epsilon/2$ to the corrupted codeword whose $j$th entry is $A(f(x), j)$ in time $poly(k, 1/\epsilon)$.

The advantage here is that for certain choices of $\epsilon$, there are efficiently list-decodable whose block length is much shorter than that for the Hadamard code, and therefore we can construct hard-core predicates for one-way permutations $f'$ whose input length is less than double that of $f$. For instance, we can concatenate Reed-Muller codes with Hadamard codes to obtain codes with message length $k$ and block length $O(k^2/\epsilon^2)$ which are efficiently list-decodable from agreement $1/2 + \epsilon$, and for which we can compute $C(x)_j$ in time $poly(k, \log 1/\epsilon)$ (to ensure that the predicate can be computed efficiently). For $\epsilon = k^{-\log k}$, this yields a construction of hard-core predicates wherein the increase in the input length for $f'$ is only $O(\log^2 k)$.

More generally, starting with a $(s, 1/s)$-one-way permutation $f : \{0,1\}^k \to \{0,1\}^k$, we can construct another $(s, 1/s)$-one-way permutation $f' : \{0,1\}^{k+O(\log s)} \to \{0,1\}^{k+O(\log s)}$ that has a $(s^{\Omega(1)}, 1/s^{\Omega(1)})$ hard-core predicate.

### 4.5.4 Pseudorandom Generators from One-way Permutations

It is not difficult to show that if $x \cdot r$ is hard to compute given $f(x), r$, then $(f(x), r, x \cdot r)$ is computationally indistinguishable from a random $(2k+1)$-bit string.[16] In particular, if $f$ is a one-way permutation, then $G : \{0,1\}^{2k} \to \{0,1\}^{2k+1}$ that sends $(x, r)$ to $(f(x), r, x \cdot r)$ is a pseudorandom generator.[17]

Once we obtain a pseudorandom generator that stretches the input by one bit, then it is possible to construct pseudorandom generators of arbitrary stretch [BM84, Yao82].

## 4.6 Goldreich-Levin and Fourier Analsys

### 4.6.1 Fourier Analysis of Boolean Functions

In the context of list-decoding of Hadamard codes, we have been looking at boolean functions $f : \{0,1\}^k \to \{0,1\}$, and linear functions $L_a : \{0,1\}^k \to \{0,1\}$, where $L_a(x) = \oplus_{i:a_i=1} x_i$ Instead, we may regard $f$ as a function $f : \{0,1\}^k \to \{1,-1\} \subseteq \mathbb{R}$ by identifying $0,1$ with $1,-1$ in the range, and $\oplus$ with $\cdot$ (multiplication over $\mathbb{R}$). In addition, consider $\chi_a : \{0,1\}^k \to \{1,-1\}$ where $\chi_a(x) = \prod_{i:a_i=1}(-1)^{x_i} = (-1)^{x \cdot a}$.

For any $f, g : \{0,1\}^k \to \mathbb{R}$, we define the dot product $f \cdot g$ to be given by $\frac{1}{2^k} \sum_x f(x)g(x)$. It is then straight-forward to verify that $\chi_a \cdot \chi_a = 1$, and $\chi_a \cdot \chi_b = 0$ for $a \neq b$. Therefore, $\{\chi_a \mid a \in \{0,1\}^k\}$ is an orthonormal basis for the set of functions $f : \{0,1\}^k \to \mathbb{R}$. This means we can write any function $f : \{0,1\}^k \to \mathbb{R}$ as:

$$f(x) = \sum_{a \in \{0,1\}^k} \hat{f}_a \chi_a(x)$$

where $\hat{f}_a = f \cdot \chi_a \in \mathbb{R}$.

---

[16]This use of hard-core predicates and one-way permutations to construct pseudorandom generators is due to Blum, Micali and Yao [BM84, Yao82].

[17]Of course we have not formally defined computational indistinguishability nor pseudorandom generator. The purpose of this short section is just to give the reader a feel for the use of hard-core predicates in cryptography. The reader is referred to the excellent and comprehensive treatment in [Gol01].

The dot product $f \cdot g$ and the Fourier coefficients $\hat{f}_a$ have a combinatorial interpretation when $f, g : \{0,1\}^k \to \{1, -1\}$ are boolean functions. Observe that

$$f \cdot g = \frac{1}{2^k} \sum_x f(x)g(x) = \Pr_x[f(x) = g(x)] - \Pr_x[f(x) \neq g(x)] = 2\Pr_x[f(x) = g(x)] - 1$$

Therefore,

$$\Pr_x[f(x) = g(x)] = \frac{1}{2} + \frac{1}{2} f \cdot g$$

and in particular,

$$\Pr_x[f(x) = L_a(x)] = \frac{1}{2} + \frac{1}{2} \hat{f}_a \quad (4)$$

Also, for any $f : \{0,1\}^k \to \mathbb{R}$, we have $f \cdot f = \sum_a \hat{f}_a^2$. In the case $f$ is a boolean function, $f \cdot f = 1$, so this yields:

$$\sum_a \hat{f}_a^2 = 1 \quad (5)$$

### 4.6.2 Learning the Fourier Coefficients of a Boolean Functions

The Goldreich-Levin theorem may be interpreted as providing an efficient algorithm that given oracle access to some boolean function $f$ and some threshold $\epsilon$, finds all linear functions $L_a$ that agrees with $f$ on at least $1/2 + \epsilon$ fraction of inputs. By (4), these linear functions correspond exactly to Fourier coefficients $\hat{f}_a$ that are at least $2\epsilon$. Furthermore, given $a$ and oracle access to $f$, estimating $\hat{f}_a$ is easy (by estimating $\Pr_x[f(x) = L_a(x)]$), so we may eliminate any extraneous coefficients that may be computed by the Goldreich-Levin algorithm.

This idea is often applied to learning classes of boolean functions $f$ for which the Fourier coefficients are concentrated on a small set $S$, that is, there is some set $S \subseteq \{0,1\}^k$ such that $|S| = poly(k)$ and $\sum_{a \in S} \hat{f}_a^2 \geq 1 - \epsilon$. Given oracle access to access to $f$, we can define a function $g$ such that $f$ and $g$ disagree on $O(\epsilon)$ fraction of inputs. Furthermore, $g$ can be efficiently computed as follows:

1. Fix $t = poly(k, 1/\epsilon)$.

2. Find a list $L$ of all $a$ such that the corresponding Fourier coefficients $\hat{f}_a$ are at least $\epsilon/t$ using the Goldreich-Levin algorithm. From (5), there are at most $O(t^2/\epsilon^2)$ such values.

3. Compute an estimate $\hat{f}'_a$ of $\hat{f}_a$ for all $a \in L$ by estimating $\Pr_x[f(x) = L_a(x)]$ using sampling.

4. Return $g = \sum_{a \in L} \hat{f}'_a \chi_a$ as an estimate for $f$.

To bound the fraction of inputs on which $f$ and $g$ disagree, we will need to bound the errors due to the omission of the small Fourier coefficients, and in the estimation of the large Fourier coefficients.

See [KM93] for this interpretation of Goldreich-Levin, and for interesting applications to learning theory.

## 4.7 More Hard-Core Predicates Using List Decoding

In this section we present the results of Akavia et al. [AGS03], which give a fresh coding-theoretic perspective to results that, previously, had only ad-hoc algebraic proofs.

The techniques of Akavia et al. give, in particular, new proofs that certain predicates are hard-core for RSA function and for exponentiation, which are the two more studied families of permutations that are conjectured to be one-way. We recall their definition below.

**Definition 9** *Given $N = pq$, $p$ and $q$ prime, choose $e$ so $gcd(e, \varphi(N)) = 1$. Then the RSA permutation $\mathbb{Z}_N \to \mathbb{Z}_N$ is*

$$\text{RSA}_{N,e}(x) = x^e \bmod N \ .$$

**Definition 10** *Given $p$ prime, $g$ a generator for $\mathbb{Z}_p^*$, the EXP isomorphism $\mathbb{Z}_{p-1} \to \mathbb{Z}_p^*$ is*

$$\text{EXP}_{p,g}(x) = g^x \bmod p \ .$$

Suppose that we have a permutation $p$ mapping $\mathbb{Z}_N$ into $\mathbb{Z}_N$, for example RSA or exponentiation, and a predicate $B : \mathbb{Z}_N \to \{0,1\}$, and that we would like to show that if $p$ is one-way that $B$ is hard-core. To prove such an implication we need to show that algorithm that computes $B(x)$ given $p(x)$ (for a fraction of $x$'s noticeably larger than $1/2$) can be transformed into an algorithm that computes $x$ given $p(x)$ (for a noticeable fraction of $x$s.)

In order to express this reduction as a coding-theoretic problem, we asume we also have a code $C : \mathbb{Z}_N \to \{0,1\}^n$ satisfying the following property.

**Definition 11 (Accessible Codes)** *$C$ is accessible with respect to $p$, $B$ if there is a probabilistic polynomial time algorithm $A$ such that*

- *For $x \in \mathbb{Z}_N$, $j \in \{1, \ldots, n\}$,*

$$A(p(x), j) = p(y) \quad s.t. \quad B(y) = C(x)_j \ .$$

- *Over the choice of random $x$ and $j$, $A(p(x), j)$ is distributed close to uniform in $\mathbb{Z}_N$.*

Intuitively, this is saying the if we are given $x$ and $j$, and we are interested in computing the $j$-th bit of the codeword $C(x)$, then we can efficiently find a string $z$ such that $B(p^{(-1)}(z))$ is equal to $C(x)_j$. Furthermore, if we are interested in finding $C(x)_j$ for a random $x$ and for a random $j$ then the string $z$ will be uniformly distributed. This means that if we have an algorithm that has a probability $1/2 + \epsilon$ of computing $B(y)$ given $p(y)$ (the probability being over a random choice of $y$), then we also have an algorithm that has probability essentially $1/2 + \epsilon$ of computing $C(x)_j$ given $x$ and $j$ (the probability being over the random choice of $x$ and $j$.) In particular, for at least an $\epsilon/2$ fraction of the $x$'s we can compute $C(x)$ on at least $1/2 + \epsilon/2$ of the entries.

We then have the following result.

**Lemma 20** *Assume $p$ is a one-way permutation, $C$ is accessible (w.r.t. p,B), and $C$ is list-decodable (from $\frac{1}{2} + \epsilon$ agreement in time $poly(\frac{1}{\epsilon}, \log N)$). Then $B$ is a hard-core predicate for $p$.*

PROOF: We only give a sketch. Assume otherwise, that on $\frac{1}{2}+\epsilon$ fraction of the $y$, we can determine $B(y)$ from $p(y)$. Over $x$ and $j$, $A(p(x), j)$ is (nearly) uniform, so there is some $\frac{\epsilon}{2}$ fraction of the $x$'s for which $\frac{1}{2} + \frac{\epsilon}{2}$ of the codeword indices $j$ are good. The list-decoding algorithm succeeds for these $x$'s. □

In order to prove that, for example, a certain predicate is hard-core for RSA, we need to define a list-decodable error-correcting that is accessible with respect to RSA and the predicate. The following family of codes will work in many instances.

**Definition 12 (Multiplication Code)** *Let $B : \mathbb{Z}_N \to \{0,1\}$ be a predicate. We define the multiplication code for $B$, $C^B : \mathbb{Z}_N \to \{0,1\}^{\mathbb{Z}_N}$ as*

$$C^B(x)_j = B(xj \bmod N) \ .$$

There are two steps now for showing that a predicate $B$ is hard-core for a permutation $p$ using this framework: show that $C^B$ is accessible for $p$, and show that $C^B$ is an error-correcting code.

Clearly, the second part of the proof depends only on $B$, and it is independent of the particular permutation $p$ that we have in mind. Perhaps surprisingly, for RSA and EXP the first part of the proof does not depend on $B$ at all.

**Lemma 21** *$\forall B$, $C^B$ is accessible with respect to RSA and EXP.*

**Definition 13** *$B : \mathbb{Z}_N \to \{0,1\}$ is a basic $t$-segment predicate if in*

$$B(0), B(1), B(2), \ldots, B(N-1), B(0) \ ,$$

*there are $\leq t$ changes of values. $B$ is $t$-segment if for some invertible $a \in \mathbb{Z}_N^*$, $B(xa)$ is basic $t$-segment.*

For example, the most-significant bit, $\mathrm{msb}(x)$, defined by

$$\mathrm{msb}(x) = \begin{cases} 1 & \text{if } x \geq \lceil N/2 \rceil \\ 0 & \text{otherwise} \end{cases}$$

is a basic 2-segment predicate. (Moreover, msb was previously known to be a hard-core predicate for RSA and EXP.)

As another example, if $N$ is odd, then the least significant bit, $\mathrm{lsb}(x) = x \bmod 2$, is a 2-segment predicate. Indeed, $\mathrm{msb}(x) = \mathrm{lsb}(2x)$ – equivalently, $\mathrm{lsb}(x) = \mathrm{msb}(x/2) = \mathrm{msb}(x\frac{N+1}{2})$. For if $x = 2y + b$ with $b \in \{0,1\}$ and $0 \leq y < \frac{N-b}{2}$, then

$$x\frac{N+1}{2} = (N+1)y + \frac{N+1}{2}b = \lceil \frac{N}{2} \rceil b + y \ .$$

Notice that EXP leaks the least significant bit (whether or not $g^x \bmod p$ is a quadratic residue). The above argument fails since $\mathrm{lsb}(2x) = 0$ in the domain $\mathbb{Z}_{p-1}$ since $p-1$ is even; 2 is not invertible.

**Theorem 22 (Main Result of [AGS03])** *Let $B : \mathbb{Z}_N \to \{0,1\}$ be a balanced, $t$-segment predicate. Then there is a list-decoding algorithm that given $t$ and $\epsilon$, and oracle access to $g : \mathbb{Z}_N \to \{0,1\}$ having agreement $\frac{1}{2} + \epsilon$ with $C^B(x)$, runs in time $\mathrm{poly}(\log N, t, 1/\epsilon)$ and outputs a list that with high probability contains $x$.*

(By "balanced," we mean that $B$ has at most a constant more, or fewer, zeros than ones. This condition could be weakened.)

We get as immediate corollaries that msb is hard-core for RSA and EXP, and, in fact, every balanced $t$-segment predicate $B$ is hard-core for RSA and EXP.

The proof of Theorem 22 has essentially four parts.

1. First, the authors consider the Fourier analysis of functions of the form $f : \mathbb{Z}_N \to \mathbb{C}$. We can think of codewords $C^B(x)$ as functions mapping $\mathbb{Z}_N$ into $\{0, 1\}$, and so, in particular, as functions mapping $\mathbb{Z}_N$ into $\{0, 1\}$. The authors show that if $B : \mathbb{Z}_N \to \{0, 1\}$ is a basic $t$-segment predicate then, for every $x$, the function $f$ correponding to the codeword $C^B(x)$ is *concentrated*, which essentially means that $f$ is well-approximated by a function that has only few non-zero coefficients.

2. Then the authors show that if $f$ is a concentrated function and $g$ is another function that agrees with $f$ on a $1/2 + \epsilon$ fraction of inputs, then there is a Fourier coefficient that is large (that is, at least some value depending on $\epsilon$ and on the "concentration" of $f$) for both $f$ and $g$.

3. The authors also show that given oracle access to a function $f : \mathbb{Z}_N \to \mathbb{C}$ and a threshold $\tau$ it is possible to efficiently find the (small) list of all the coefficients of $f$ that are larger than $\tau$.

4. Finally, for every fixed $t$-segment predicate $B$, there is an algorithm that given a Fourier coefficient finds the (small) list of all the strings $x$ such that that coefficient is large for the function corresponding to $C^B(x)$

Having proved all these results, the list-decoding algorithm is as follows: we first find (using part 3 above) all the large Fourier coefficients of $g$, where "large" means larger than a threshold that is polynomial in $\epsilon$, $1/t$ and $1/\log N$. Then, for each of these coefficients, we find (using part 4) all the strings $x$ such that $C^B(x)$ is large in that coefficient.

## 4.8 Notes and References

Hard-core predicates appear in the work of Blum and Micali [BM84] and of Goldwasser and Micali [GM84], and they were defined in a more general setting by Yao [Yao82], who showed that every one-way permutation can be modified to have a hard-core predicate. Levin [Lev87] gives a different proof that uses error-correcting codes. Goldreich and Levin [GL89] give a more efficient construction of hard-core predicates. As previously disccused, the Goldreich-Levin algorithm can be seen as a sub-linear time list-decoding procedure for the Hadamard codes. Goldreich and others [GRS00] give a list-decoding algorithm for the Reed-Muller codes that runs in sub-linear time for certain ranges of the parameters. The algorithm is required to output a list of messages, rather than a list of implicit representations, and so the linear time cannot be sublinear in $k$. (Although it is sub-linear in $n$ if $n$ is much larger than $k$.) A general connection between list-decoding and hard-core predicates was recognized in unpublished work in mid 1990s by Impagliazzo and Sudan. Kushilevitz and Mansour [KM93] recognized the connection between Goldreich-Levin and Fourier analysis, and its applications to learning.

Prior to the work of Akavia and others [AGS03], hard-core predicates for specific algebraic one-way permutations were proved with ad-hoc techniques. The work of Akavia and others [AGS03] combines learning, list-decoding, Fourier analysis and hard-core predicates in a very surprising generalization of the techniques of [GL89].

Sudan and others [STV01] present sublinear time list decoding algorithms for Reed-Muller codes, with applications to worst-case to average-case complexity. The connection between coding theory and worst-case to average-case connection is further discussed in [TV02, Vio03, Tre03].

## 5 Locally Testable Codes

In this section we consider codes with sub-linear time error-detection algorithms. We look for algorithm that are able to distinguish valid codewords from strings that are "far" in Hamming distance from all codewords.

**Definition 14 (Locally Testable Code (LTC))** *A code $C : \Sigma^k \to \Sigma^n$ is $(q, \delta, p)$-locally testable if there is an oracle algorithm A of query complexity q such that*

- *For every message $x$, $\Pr[A^{C(x)}\text{accepts}] = 1$*

- *For every string $y$ that has distance at least $\delta n$ from all codewords of $C$, $\Pr[A^y \text{accepts}] \leq p$.*

This notion was introduced by Rubinfeld and Sudan [RS96] and by Friedl and Sudan [FS95], and it also appears (under the name of "probabilistically checkable" codes) in Arora's PhD thesis [Aro94] and (under the name "checkable" codes) in Spielman's PhD thesis [Spi95].

**Remark 2** *We make a few remarks about the definition.*

- *A stronger condition is to say that the code is $(c, q)$-locally checkable if there is an algorithm A of query complexity q that satisfies the first part of the above definition, and such that if y is a string that is at distance at least d from all codewords then $\Pr[A^y \text{accepts}] \leq 1 - cd/n$. This means that the algorithm has, as in the above definition, a constant probability of rejecting strings that are at more than a certain constant minimum distance from all codewords, but also that it has a non-zero probability of rejecting any non-codeword, and that the rejecting probability grows linearly with the distance of the code. Many positive results about locally checkable codes prove that this stronger codition is satisfied.*

- *We gave a one-sided error definition. A two-sided error definition could also be given and it would also make sense.*

While the notion of locally testable codes was introduced only around 1994, local testers for the Hadamard code [BLR93] and for the Reed-Muller code [GLR+91] had been studied in the context of program self-testing, and had found their most powerful application in the construction of probabilistically checkable proofs.

### 5.1 Probabilistically Checkable Proofs

**Definition 15 (Probabilistically Checkable Proofs)** *Let L be a language and V be a probabilistic polynomial time oracle machine. We say that V is a $(q(n), r(n))$-restricted PCP verifier for L if the following conditions hold:*

- On input $x$ of length $n$, and for every oracle $\pi$, $V^\pi(x)$ makes at most $q(n)$ oracle queries and tosses at most $r(n)$ random bits.

- If $x \in L$ then there is a string $\pi$ such that $\mathbf{Pr}[V^\pi(x)\text{accepts}] = 1$.

- If $x \notin L$ then for every string $\pi$, $\mathbf{Pr}[V^\pi(x)\text{accepts}] \leq 1/2$.

We denote by $\text{PCP}[r(n), q(n)]$ the class of languages that have $(r(n), q(n))$-restricted verifiers.

We think of $\pi$ as being a "proof" that $x \in L$. Such a proof needs only be of length $2^{r+q}$, because this is the maximum number of distinct oracle queries that the machine can make. If $r + q = O(\log n)$ then the proof $\pi$ is of polynomial length, and the verification process of $V$ can be derandomized by running through all possible $r$, so every language in $\text{PCP}[O(\log n), O(\log n)]$ is also in NP, and the proof $\pi$ can be thought of as an NP witness for $x$.

The stunning result about PCP is that every NP witness can be put in such a form that it can be checked with high confidence in constant time.

**Theorem 23 (PCP Theorem [AS98, ALM$^+$98])** $\text{NP} = \text{PCP}[O(\log n), O(1)]$.

Rather than trying to survey constructions, applications and ideas in the area of PCP[18], I will discuss a recent development that will hopefully lead to a simpler proof of the PCP theorem and possibly to locally testable codes and PCPs of optimal length.

## 5.2 PCPs of Proximity

Consider the following definition.

**Definition 16 (PCP of Proximity [BSGH$^+$04, DR04])** *An $(r(n), q(n))$-restricted PCP of $\delta$-Proximity for an NP relation $R$ is a probabilistic polynomial time oracle algorithm $A$ such that*

- *If $(x, y) \in R$, then there is a $\pi$ such that $\mathbf{Pr}[A^{y,\pi}(x)\text{accepts}] = 1$.*

- *If $\mathbf{Pr}[A^{y,\pi}(x)\text{accepts}] \leq 1/2$. then $y$ is $\delta$-close to a $y'$ such that $(x, y') \in R$.*

A PCP of proximity is the same as a standard PCP, except for the fact the proof is composed of two part: a standard witness $y$ and a possibly very complex component $\pi$. When the verifier accepts with high probability, then not only it has confidence that a witness exists for input $x$, but actually it has confidence that $y$ itself is close to a witness.

The definition becomes more clear, perhaps, when specialized to a particular NP problem, say, circuit satisfiability.

**Definition 17 (Assignment Tester)** *An assignment tester is a PCP of proximity for the circuit satisfiability problem. Formally, a $(\delta, r(n), q(n))$ Assignment Tester is a probabilistic polynomial time oracle algorithm $A$ such that*

- *If $C$ is a circuit and $a$ is an assignment such that $C(a) = 1$, then there is a $\pi$ such that $\mathbf{Pr}[A^{a,\pi}(C)\text{accepts}] = 1$.*

- *If $\mathbf{Pr}[A^{a,\pi}(C)\text{accepts}] \leq 1/2$. then $a$ is $\delta$-close to a $a'$ such that $C(a') = 1$.*

---
[18]A task covered by many survey papers, although my favorite introduction to the area is the introduction of a research paper by Bellare et al. [BGS98].

## 5.3 Relations between PCPs of Proximity, PCP, and Locally Testable Codes

We have already observed that PCP of Proximity is only a stronger algorithm than a PCP verifier. The randomness, query complexity and completeness constraints are the same. The soundness constraint in the definition of PCPP not only implies that $x$ is in the language, but it also implies the stronger property that there is a witness that $x \in L$ which is close to the initial segment of the oracle proof.

It is less trivial, but still simple, to get a LTC from an AT. Let $\mathcal{C}$ be an error-correcting code , and let $C$ be a circuit that checks whether a given string is a codeword of the code, let $V$ be an assignment tester for $C$. Then for every codeword $\mathcal{C}(x)$ there is a proof $\pi_x$ such that $\mathcal{C}(x), \pi(x)$ is accepted with probability 1 by the assigment tester $V$. Suppose now that $(y, w)$ is accepted with probability higher than $1/2$ by $V$: then $y$ is close to a valid codeword $\mathcal{C}(x)$. If we could argue that $(y, w)$ is close to $(\mathcal{C}(x), \pi_x)$, then we would have shown that the mapping $x \to (\mathcal{C}(x), \pi_x)$ is a good error-correcting code. Unfortunatelty, if the length of the proof is large compared with the length of the assignment, then it is not possible to conclude that $(y, w)$ is close to $(\mathcal{C}(x), \pi_x)$ just because $y$ is close to $\mathcal{C}(x)$. This is indeed a problem because in all known constructions the length of the proof is super-linear in the length of the assignment.

This problem is resolved by considering the mapping $x \to (\bar{\mathcal{C}}(x), \pi_x)$ where $\bar{\mathcal{C}}(x)$ is a sequence of several identical copies of $\mathcal{C}(x)$, enough copies so that the total length is equal to about $O(1/\delta)$ times the length of $y$.

Our tester is given a string $(y_1, \ldots, y_k, w)$, where the $y_i$ are supposed to be identical copies of the same codeword $\mathcal{C}(x)$, and $w$ is supposed to be the proof $\pi_x$ that $\mathcal{C}(x)$ is a valid codeword. The tester first checks that strings $y_i$ are approximately equal. This is done by repeatedly picking a pair $i, j$ and then comparing the strings $y_i, y_j$ in a random position. Then the tester simulates the algorithm of the assignment tester. Each query of the assignment tester to the assignment part of the oracle is randomly routed to one of the strings $y_i$. Clearly a valid codeword is accepted with probability 1. The analysis is completed by arguing that if the test accepts with high probability (say, larger than $3/4$), then the strings $y_i$ are approximately all equal and that a "majority" decoding into a single string y is such that $y, \mathcal{C}(x)$ would have been accepted by the assignment tester with probability at least $1/2$. So $y$ is close to a valid codeword, and $y_1, \ldots, y_k$ is close to a valid repetition of a valid codeword. Details are in the full version of [BSGH+04].

From [SS96, Spi96], we know that there are error-correcting codes $\mathcal{C} : \{0, 1\}^k \to \{0, 1\}^n$ with constant relative minimum distance, for which $n = O(k)$ and such that there is a circuit of size $O(k)$ that recognizes valid codewords. If there were assignment testers with constant query complexity and linear proof length, then the above argument would show the existence of an LTC with block length $O(k)$, that is, an asymptotically optimal LTC.

Assignment testers with linear proof length are not known, and the best known construction is as follows.

**Theorem 24 ([BSGH+04])** *For every constant $\epsilon$ there is a constant $q$ such that there is an Assignment Tester of query complexity $q$ that, for a circuit of size $n$, uses $O(\log n)$ randomness and expects a proof of length $2^{O((\log n)^\epsilon)}$.*

We remind the reader that results from [KT00] imply that a locally decodable code with a decoder having constant query complexity cannot have encoding length $k^{1+o(1)}$, so the above result already shows a separation between the rate of LTCs versus LDCs with comparable parameters.

The main open question is clearly

**Open Question 4** *Are there LTCs with constant query complexity and constant rate?*

Similarly, we could ask if there are assignment testers with proofs of linear length. If there were an assignment tester with logarithmic randomness, constant query complexity and a proof of linear length, then there would a randomized reduction from, say, 3SAT to the Max CUT problem. Starting from a 3SAT instance with $n$ variables and $O(n)$ clauses, the reduction would produce an instance of Max CUT with $N = O(n)$ nodes and $M = O(n)$. For some fixed constants $p$ and $\epsilon$, a satisfiable 3SAT instance would produce a graph where the size of the maximum cut is at least $p$; an unsatisfiable instance of 3SAT would produce a graph where the size of the maximum cut is at most $p(1 - \epsilon)$. Consider now the following question.

**Open Question 5** *Is it possible to approximate the Max CUT problem in bounded-degree graphs to within a factor $1 + o(1)$ in time $2^{o(n)}$?*

A positive answer would imply that an assignment tester like the one discussed above could be used to get a $2^{o(n)}$ algorithm for 3SAT, a conclusion that is typically considered to be unlikely. A positive answer to Question 5 could then be taken as evidence that assignment testers need proofs of super-linear length.

## 5.4 Notes and References

The PCP Theorem was the culmination of a long line of collaborative work, that is difficult to summarize.

In telling this story, one typically starts from the introduction of the model of "interactive proof systems" due independently to Goldwasser, Micali and Rackoff [GMR89] and to Babai [Bab85]. In this model, a probabilistic verifier *interacts* with a prover, as opposed to receiving a fixed proof and checking its validity. The work of Goldwasser, Micali and Rackoff [GMR89] also introduces the notion of "*zero-knowledge* proof system," which later became a fundamental primitive in the construction of cryptographic primitives. A fundamental result by Goldreich, Micali and Wigderson [GMW91] shows that every problem in NP has a zero-knowledge proof system, assuming that a certain cryptographic assumption is true. Ben-Or et al. [BOGKW88] considered a model of zero-knowledge where the verifier can interact with two (or, more generally, several) provers, who are all computationally unbounded *but* unable to communicate with each other once the protocol starts. The contribution of [BOGKW88] was to show that every problem in NP has a zero-knowledge proof system in this model, without cryptographic assumption. The model of multi-prover interactive proof (without the zero-knowledge requirement) was further studied by Fortnow, Rompel and Sipser [FRS88]. They show that the class of languages admitting such proof systems has the following equivalent characterization: it can be seen as the class of languages that admit exponentially long proofs of membership that can be checked in polynomial time by a randomized verifier (with bounded error probability). This class is clearly contained in NEXP, where NEXP is the class of decision problems that admit exponentially long proofs that can be checked in exponential time in the length of the input (but, without loss of generality, in polynomial time in the length of the proof itself).

Initially, it was conjectured that MIP was only a small extension of NP, and that coNP $\not\subseteq$ MIP. Shortly after Shamir's proof that IP = PSPACE [Sha92], Babai, Fortnow and Lund [BFL91] showed that MIP = NEXP. This is a truly impressive result: it says that for every language that admits exponentially long proofs, such proofs can be encoded in such a way that a polynomial-time randomized verifier can check them. The verifier will accept correct proofs with probability 1, and "proofs" of incorrect statements with probability $\leq 1/2$ (or, equivalently, with probability exponentially small in the length of the input). So the verifier becomes convinced of the validity of the proof even if it only looks at a negligible part of the proof itself.

It is natural to ask whether polynomially long proofs can be checked in polylogarithmic time. This question has to be phrased carefully, since a polylogarithmic time verifier cannot even read the instance, which makes it impossible to verify a proof for it. However if both the instance and the proof are encoded in a proper (efficiently computable) way, then Babai et al. show that polylogarithmic time verification is possible [BFLS91]. A variant of this result was also proved by Feige et al. [FGL+91]: they show that NP-proofs have a quasi-polynomial length encoding (i.e. an encoding of length $n^{O(\log \log n)}$) such that a polynomial-time verifier can verify the correctness of the proof in polynomial time by using $O(\log n \log \log n)$ random bits and reading $O(\log n \log \log n)$ bits of the proof. The main result of Feige et al. [FGL+91] was to show a connection between the computational power of such a model and the hardness of approximating the Max Clique problem.[19] The result of Feige et al. [FGL+91] can be written as $\text{NP} \subseteq \text{PCP}[O(\log n \log \log n), O(\log n \log \log n)]$.

Arora and Safra [AS98] introduced several new ideas to improve on [FGL+91], and proved that $\text{NP} = \text{PCP}[O(\log n), O(\sqrt{\log n})]$. The main contribution of Arora and Safra is the idea of "composing" proof systems together. The next step was to realize that the reduction from PCP to Max Clique was not an isolated connection betweeb proof checking and approximability. Sudan and Szegedy (as credited in [ALM+98]) discovered that the computations of a $(O(\log n), O(1))$-restricted verifier can be encoded as instances of the Max 3SAT problem. Then, using the web of reductions between optimization problems initiated by Papadimitriou and Yannakakis [PY91], this also implies that the strength of $(O(\log n), O(1))$-restricted verifiers implies the hardness of approximating several important problems including the Traveling Salesman Problem and the Steiner Minimal Tree problems in metric spaces. This was a strong motivation to prove the PCP Theorem, that came only a few months after the initial circulation of the paper of Arora and Safra [AS98].

Further work on strengthening relation bewteen query complexity, error probability and other parameters, and improve hardness of approximation. In this section we have concentrated on the problem of proof length and construction of locally testable codes, a question that has received relatively less attention until recently.

Locally testable codes were discussed in many places, including [Aro94, Spi95, FS95]. Constructions of short PCPs were first presented by Polishuck and Spielman [PS94]. Friedl and Sudan [FS95] construct both short PCPs and good locally testable codes, and further improvements are due, more recently, to Harsha and Sudan [HS00].

Goldreich and Sudan [GS02] give a nearly-linear length construction of locally testable codes and PCPs. The result of [GS02] is based on a probabilistic construction, and so the codes are not computable in polynomial time, although they can be computed by polynomial size circuits. Similarly, the verifier in their PCP construction can be realized by a polynomial size circuit but not by a uniform machine. The results of Goldreich and Sudan have been improved and made explicit in [BSSVW03, BSGH+04]. There is no lower bound for locally testable codes, except the one in [BSGS03] for a very special case.

Dinur and Reingold have recently made considerable progress towards a simpler and more "combinatorial" proof of the PCP theorem [DR04], a direction of work on which there is essentially no previous result, except for interesting work by Goldreich and Safra [GS00a].

---

[19] A clique in a graph is a subset of vertices that are all pairwise adjacent. The Max Clique problem is, given a graph, to find the largest clique.

## Acknowledgements

Most of my understanding of coding theory comes from Madhu Sudan. Working with Salil Vadhan on various research project has clarified connections to complexity. Irit Dinur has been extremely patient in explaining to me the results of [DR04]. Thanks to the anonymous referee for providing several additional references and several helpful suggestions.

In Fall 2003 I taught a graduate class on applications of coding theory to computational complexity, I thank the students who took the class for their enthusiasm and participation.

# A Appendix

## A.1 The Berlekamp-Welch Algorithm

Consider the following algorithm:

1. If there is a polynomial $p$ such that $p(x_i) = y_i$ for all $i = 1, \ldots, n$, output $p$. Otherwise:

2. Find polynomials $E(x)$ and $N(x)$ such that

   (a) $E$ is not identically zero;
   
   (b) $E(x)$ has degree at most $e$ and $N(x)$ has degree at most $e + k - 1$;
   
   (c) For every $i = 1, \ldots, n$, $N(x_i) = E(x_i) \cdot y_i$.

3. Output $N(x)/E(x)$, or output `error` if $N(x)$ is not a multiple of $E(x)$.

We claim that the algorithm can be implemented to run in $O(n^3)$ time and that it correctly finds the unique solution.

Let $p$ be the unique solution, and let $I = \{i : p(x_i) \neq y_i\}$. If $I$ is empty, then the algorithm finds $p$ in step (1). We want to show that, if $I$ is not empty, then steps (2) and (3) can be implemented in $O(n^3)$ time and that the algorithm finds $p$ in step (3).

Regarding efficiency, polynomial division can be realized in almost linear time, so we only need to worry about step (2). We can write $E(x) = \sum_{i=0}^{e} a_i x^i$ and $N(x) = \sum_{i=0}^{e+k-1} b_i x^i$, and see the problem of realizing step (2) of the algorithm as the problem of finding coefficients $a_i$ and $b_i$ such that the constraints $N(x_i) = E(x_i) y_i$ are satisfied. Such constraints are *linear* in $a_i$ and $b_i$, and so, if the set of constraints has a non-zero solution, then a non-zero solution can be found in cubic time using Gaussian elimination.

To see that a non-zero solution exists, let us define the polynomials $E(x) = \prod_{i \in I}(x - x_i)$ and $N(x) = E(x) \cdot p(x)$. Then by definition the degree of $E$ is at most $e$, and the degree of $N$ is at most $k - 1 + e$. Furthermore, if $i \in I$ we have $E(x_i) = 0$ and $N(x_i) = 0$, so that $N(x_i) = E(x_i) y_i = 0$; if $i \notin I$ we have $N(x_i) = E(x_i) p(x_i) = E(x_i) y_i$, and so all the constraints are satisfied. Finally, $I$ is not empty (otherwise we would have found $p$ at step (1) of the algorithm) and so $E$ is not the all-zero polynomial.

Regarding correctness, let $E, N$ be the polynomials defined above, and let $E', N'$ be the solution found by the algorithm in step (2), we want to argue that $N(x)/E(x) = N'(x)/E'(x)$, which is the same as $N(x)E'(x) = N'(x)E(x)$. The polynomials $N(x)E'(x)$ and $N'(x)E(x)$ have degree at most $2e + k - 1 < n$, and so, to show that they are equal, it is enough to show that they agree in $n$ inputs. This is easily verifier because, for every $i = 1, \ldots, n$, we have

$$N(x_i)E'(x_i) = y_i E(x_i) E'(x_i) = N'(x_i) E(x_i)$$

## A.2 List-Decoding of the Reed-Solomon Code

### A.2.1 A Geometric Perspective

For the purpose of this algorithm, we will want to describe the $n$ given points using low-degree planar curves that pass through them; that is, we consider curves $\{(x, y) : Q(x, y) = 0\}$ where $Q(x, y)$ is a low-degree polynomial. Note that we are not restricted to curves with degree one in $y$;

in particular, we may describe points on a circle centered at $(0,0)$ with the equation $x^2 + y^2 - 1 = 0$. Other examples of point sets that may be described using low-dimensional curves are lines, and unions of lines and circles.

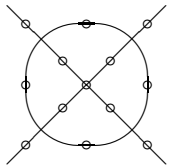

In the example on the left, we have a set of 13 points that lie on the union of a circle and two lines. Suppose the point in the center is $(0,0)$. Then, the set of points lie on the curve described by: $(x^2 + y^2 - 1)(x - y)(x + y) = 0$.

### A.2.2 A Simple List-Decoding Algorithm

For the list-decoding problem, the intuition is that if $p$ is a polynomial with large agreement, then the curve $y - p(x) = 0$ passes through many of the given points. Therefore, what we will do in the list-decoding algorithm is to first find a low-degree polynomial $Q$ passing through all of the given points, and then show that all low-degree curves that pass through many of the given points divides $Q$. This reduces the list-decoding problem to factorizing a bivariate polynomial over a finite field, for which efficient algorithms do exist.

**Algorithm** LIST-DECODE-RS
**Given:** $n$ distinct points $(x_1, y_1), \ldots, (x_n, y_n)$ in $\mathbb{F}_q^2$.

1. Find $Q(x, y)$ such that
    - $Q$ has low degree: $d_x - 1$ in $x$, $d_y - 1$ in $y$
    - $Q(x_i, y_i) = 0$ for all $i = 1, 2, \ldots, n$
    - $Q \not\equiv 0$.

2. Factor $Q(x, y)$. For every factor of the form $y - p(x)$, if $p$ is a feasible solution, output $p$.

There are two problems that we need to address:

1. Does there exist a low-degree polynomial $Q$ that pass through all the given points, and if so, how can we find one efficiently?

2. Must every low-degree polynomial that pass through many of the given points divide $Q$? For instance, taking $t = 3$ for concreteness; it seems conceivable that we have a polynomial $R(x, y)$ quadratic in $y$ that passes through 6 of the given points that lie on $y - p_1(x), y - p_2(x)$ for two low-degree polynomials $p_1, p_2$, and that $R(x, y)$ divides $Q$, but neither $y - p_1(x)$ nor $y - p_2(x)$ does.

### A.2.3 Finding $Q$

First, we address the problem of finding $Q$. We may write

$$Q(x, y) = \sum_{i=0,\ldots,d_x-1;\, j=0,\ldots,d_y-1} c_{ij} x^i y^j$$

Now, the problem reduces to finding the $d_x d_y$ coefficients of $Q$: $c_{ij}, i = 0, 1, \ldots, d_x - 1; j = 0, \ldots, d_y - 1$. Observe that the requirement $Q(x_i, y_i) = 0$ is equivalent to a system of linear constraints on the coefficients $\{c_{ij}\}$. Furthermore, this is a homogeneous system, so it will always have the all 0's solution, corresponding to $Q \equiv 0$. On the other hand, if $d_x d_y > n$, that is, the number of variables is more than the number of linear constraints, then we can always efficiently find a non-zero solution to the linear system that yields a non-zero $Q$ that passing through all the $n$ points.

### A.2.4 Proof of Correctness

Next, we will have to show that every polynomial $p$ with large agreement with the points $(x_1, y_1), \ldots, (x_n, y_n)$ is a factor of $Q$. More precisely, we are told that:

1. $p(x)$ is a degree $k - 1$ polynomial such that $y - p(x)$ is zero in at least $t$ of the points.

2. $Q(x, y)$ has degree $d_x - 1$ in $x$ and $d_y - 1$ in $y$ and passes through all the points (that is, $Q(x_i, y_i) = 0$ for $i = 1, 2, \ldots, n$).

3. There are $\geq t$ points $(x_i, y_i)$ such that $Q(x_i, y_i) = y_i - p(x_i) = 0$.

For simplicity, we can rewrite these conditions assuming that we are choosing $n$ points on the curve $Q(x, y) = 0$, which yields the following statement:

**Proposition 25** *Suppose that*

1. *$Q(x, y)$ is bivariate polynomial in $x, y$ with degree $d_x - 1$ in $x$ and $d_y - 1$ in $y$.*

2. *$p(x)$ is a degree $k - 1$ polynomial in $x$.*

3. *There are $\geq t$ points $(x_i, y_i)$ such that $Q(x_i, y_i) = y_i - p(x_i) = 0$.*

4. *$t > (d_x - 1) + (k - 1)(d_y - 1)$.*

*Then $y - p(x)$ divides $Q(x, y)$.*

This proposition is a special case of Bezout's Theorem, that says that any two curves that share lots of points in common must share a common factor. Here, $y - p(x)$ is irreducible (over polynomials in $y$ with coefficients from $\mathbb{F}_q[x]$), so it divides $Q(x, y)$. A simple proof of this special case is shown below.

It is also important to note that we only require that the points $(x_1, y_1), \ldots, (x_n, y_n)$ be distinct, and not that $x_1, \ldots, x_n$ be distinct, as in the case for list-decoding Reed-Solomon codes. This allows the list-decoding procedure to be used in a more general setting, as we shall see later.

PROOF: View $Q$ is a univariate polynomial in $y$ whose coefficients are univariate polynomials in $x$:

$$q(y) = q_0(x) + y q_1(x) + \ldots + y^{d_y - 1} q_{d_y - 1}(x)$$

Recall the Factor Theorem for polynomials: $\beta$ is such that $q(\beta) = 0$ iff $y - \beta$ divides $q(y)$. This tells us that $p(x)$ is such that $q(p(x)) \equiv 0$ iff $y - p(x)$ divides $Q(x, y)$. Therefore, to show $y - p(x)$ divides $Q(x, y)$, it suffices to show that $Q(x, p(x))$ is the zero polynomial.

From condition 3, we know that $Q(x_i, p(x_i)) = 0$ for at least $t$ distinct values of the $x_i$'s. On the other hand, $Q(x, p(x))$ as a univariate polynomial in $x$ can be written as:

$$Q(x, p(x)) = q_0(x) + p(x)q_1(x) + \ldots + p(x)^{d_y-1}q_{d_y-1}(x)$$

and has degree at most $(d_x - 1) + (k - 1)(d_y - 1)$. Therefore, if $t > (d_x - 1) + (k - 1)(d_y - 1)$, then $Q(x, p(x)) \equiv 0$ and $y - p(x)$ divides $Q(x, y)$. $\square$

### A.2.5 Fixing the Parameters

We are now ready to fix the parameters $d_x, d_y$. Recall that we require that:

1. $d_x d_y > n$, so that we have sufficient variables in the linear system for finding $Q$;

2. $t > d_x + k d_y$, to ensure that every polynomial with large agreement is a factor of $Q$.

We want to maximize $t$ under both constraints, and that is optimized by setting $d_x = \sqrt{kn}$ and $d_y = \sqrt{n/k}$, so $d_x + kd_y = 2\sqrt{kn}$. As a polynomial in $y$, $Q$ has degree $d_y$ and therefore at most $d_y$ factors. Hence, there are at most $d_y = \sqrt{n/k}$ polynomials in the list. This yields the following results:

**Theorem 26** *Given a list of $n$ points $(x_1, y_1), \ldots, (x_n, y_n)$ in $\mathbb{F}_q^2$, we can efficiently find a list of all polynomials $p(x)$ of degree at most $k - 1$ that pass through at least $t$ of these $n$ points, as long as $t > 2\sqrt{nk}$. Furthermore, the list has size at most $\sqrt{n/k}$.*

**Theorem 27** *For every $\epsilon > 0$, and for all sufficiently large $n$, there exist:*

1. *A $[n, \epsilon n, (1-\epsilon)n]_n$ Reed-Solomon code, such that we can efficiently list-decode from agreement in at least $2\sqrt{\epsilon}n$ locations, and size of the list is at most $\sqrt{1/\epsilon}$.*

2. *A $[n, \epsilon^2 n/4, (1 - \epsilon^2/4)n]_n$ Reed-Solomon code such that we can efficiently list-decode from agreement in at least $\epsilon n$ locations, and the size of the list is at most $2/\epsilon$.*

### A.2.6 Increasing the List-Decoding Radius

Observe that in the proof of correctness, we only require that $Q(x, p(x))$ has degree less than $t$ in $x$. Therefore, it suffices that for all monomials $x^i y^j$ in $Q(x, y)$, we have $i + kj < t$ (instead of the more restrictive constraint that $i < t/2$ and $j < t/2k$). This means that we may consider any $Q(x, y)$ of the form:

$$Q(x, y) = \sum_{i+kj<t} c_{ij} x^i y^j$$

Therefore, the number of coefficients (and thus the number of variables in the linear system) is given by:

$$|\{(i,j) : i + kj < t\}| = \overbrace{t + (t-k) + (t-2k) + \ldots + (t - \frac{t}{k} \cdot k)}^{t/k} = \frac{t}{k} \cdot \frac{1}{2}(t+0) = \frac{t^2}{2k}$$

(instead of $t/2 \cdot t/2k = \frac{t^2}{4k}$ if we consider only $i < t/2$ and $j < t/2k$.) To ensure that the linear system $\{Q(x_i, y_i) = 0 \mid i = 1, 2, \ldots, n\}$ is under-determined, we need $\frac{t^2}{2k} > n$, or equivalently, $t > \sqrt{2kn}$. For such $t$, it suffices to consider $Q$ of the form:

$$Q(x, y) = \sum_{i+kj < t \mid j \leq \sqrt{2n/k}} c_{ij} x^i y^j$$

This allows us to place an upper bound of $\sqrt{2n/k}$ on the size of list (instead of the crude bound $t/k$).

**Theorem 28** *Given a list of $n$ points $(x_1, y_1), \ldots, (x_n, y_n)$ in $\mathbb{F}_q^2$, we can efficiently find a list of all polynomials $p(x)$ of degree at most $k - 1$ that pass through at least $t$ of these $n$ points, as long as $t > \sqrt{2nk}$. Furthermore, the list has size at most $\sqrt{2n/k}$.*